\documentclass[twocolumn,showpacs,aps,prd,amsmath,amssymb,nobibnotes,floatfix]{revtex4-1}
\newcommand{\beq}{\begin{equation}}
\newcommand{\eeq}{\end{equation}}
\newcommand{\beqn}{\begin{eqnarray}}
\newcommand{\eeqn}{\end{eqnarray}}

\newcommand{\apjl}{Astrophys.\ J.\ Lett.}

\newcommand{\llabel}[1]{\label{#1}}              % DO NOT show equation label
\newcommand{\labeq}[2]{ \begin{equation} \llabel{#1}
{#2}
\end{equation}}

\newcommand{\espl}{ \end{split} }
\newcommand{\bspl}{ \begin{split} }

\usepackage{wasysym}
\usepackage{graphicx}
\usepackage{epsf}
\usepackage{graphics,epsfig,placeins,subfigure,wrapfig}

\begin{document}

%%%%%%%%%%%%%%%%%%%%%%
%%%   Title
%%%%%%%%%%%%%%%%%%%%%%

\title{Pulsar spin-down luminosity: Simulations in general relativity}
\author{Milton Ruiz$^{1}$}
\email{ruizm@illinois.edu}

\author{Vasileios Paschalidis$^{1}$}
\email{vpaschal@illinois.edu}

\author{Stuart L. Shapiro$^{1,2}$}
\email{slshapir@illinois.edu}

\affiliation{$^{1}$Department of Physics, University of  Illinois at Urbana-Champaign,
Urbana, IL 61801\\
$^{2}$Department of Astronomy \& NCSA, University of  Illinois at Urbana-Champaign,
Urbana, IL 61801}

%%%%%%%%%%%%%%%
%%% Abstract
%%%%%%%%%%%%%%%

\begin{abstract}
 Adopting our new method for matching general relativistic, ideal
 magnetohydrodynamics to its force-free limit, we perform the first
 systematic simulations of force-free pulsar magnetospheres in general
 relativity. We endow the neutron star with a general relativistic
 dipole magnetic field, model the interior with ideal
 magnetohydrodynamics, and adopt force-free electrodynamics in the
 exterior. Comparing  the spin-down luminosity to its corresponding
 Minkowski value, we find that general relativistic effects give rise
 to a modest enhancement: the maximum enhancement for $n=1$ polytropes
 is $\sim 23\%$. Evolving a rapidly rotating $n=0.5$ polytrope we find
 an even greater enhancement of $\sim 35\%$. Using our simulation data, we derive
 fitting formulas for the pulsar spin-down luminosity as a function of
 the neutron star compaction, angular speed, and dipole magnetic
 moment. We expect stiffer equations of state and more rapidly
 spinning neutron stars to lead to even larger enhancements in the
 spin-down luminosity.

\end{abstract}

%%%%%%%%%%%%%%%%
%%%   PACS   %%%
%%%%%%%%%%%%%%%%


\maketitle
%\tableofcontents

%%%%%%%%%%%%%%%%%%%%
%%% Introduction
%%%%%%%%%%%%%%%%%%%%

\section{Introduction}
\label{sec:introduction}

Pulsars are believed to be rapidly rotating neutron stars (NSs) that
lose their rotational kinetic energy primarily due to emission of
electromagnetic radiation (see e.g. ~\cite{ShaTeu83a}). Pulsars are
extremely accurate clocks that can be used to probe fundamental
physics, such as the nuclear equation of state (EOS), and theories
of gravity (see e.g. \cite{lrr-2008-9}). They can even function as
detectors of gravitational waves~\cite{0264-9381-27-8-084013}.

Pulsars are detectable from radio to gamma-ray frequencies. However,
only a small number of pulsars is observable in the visible bands
\cite{2002nsps.conf...44S}. Fermi-LAT technology
\cite{Atwood:2009ez} has allowed the recent discovery of numerous
gamma-ray pulsars. To date there are over 2300 known pulsars
\cite{2005AJ....129.1993M} of which over $130$ are gamma-ray pulsars
\cite{FERMILATGPULSARS,TheFermi-LAT:2013ssa}: the vast majority of
pulsars are radio pulsars. The most slowly rotating  pulsar is
PSR-J2144-3933~\cite{2005ApJ...627..397Z}, with a period of 9.43 s.
The most rapidly rotating pulsar is PSR-J1748-2446ad
\cite{2006Sci...311.1901H}, with a period of 1.395 ms. Hence pulsar
periods cover almost 4 orders of magnitude.

A complete theory that can explain both the radio and the gamma-ray
emission remains elusive.  Nevertheless, our understanding of
pulsar physics has been drastically improved in recent years as a
result of computational simulations that model the global pulsar
magnetosphere.

Rapid rotation in pulsars induces strong electric fields capable of
stripping matter off the NS surface and eventually populating the
exterior with tenuous plasma \cite{Goldreich:1969sb}. Motivated by
this result, most global studies of pulsar magnetospheres adopt the
force-free (FFE) limit of magnetohydrodynamics (MHD), which is valid 
in such environments. Moreover, to simplify the analysis further, 
most studies consider the case where the NS dipole magnetic moment is aligned or
antialigned with its angular momentum. This simple model has been
successful in producing the main features of the pulsar magnetosphere of
an aligned rotator: (i) a near or closed zone, which corotates with
the star, and in which magnetic field lines return to the stellar
surface; (ii) a far or open zone, which extends beyond the
light cylinder radius $R_{\rm LC}$, and in which the magnetic lines
are open and extend to infinity; (iii) a Y point at the
location of the light cylinder, at which the magnetic field lines first 
open; and (iv) an equatorial current sheet. All these
features are predictions of the so-called pulsar equation
\cite{Michel73,Scharleman73}.

The first successful numerical solution of the pulsar equation was
presented in~\cite{Contopoulos:99}, which was later followed by
numerous studies ~\cite{2002astro.ph.11141K,McKinney:2006sd,
  2006ApJ...648L..51S,Gruzinov:2008gb,Gruzinov:2013pva,
  Tchekhovskoy:2012hm,Uzdensky:2012tf} that probed the global features
of an aligned rotator in flat spacetime and reinforced the global
picture outlined above. These studies did not include the magnetized
NS interior and modeled the effects of rotation through a boundary
condition on the spherical stellar surface. Simulations of force-free
magnetospheres have produced important results, such as a proof of
existence of a stationary force-free magnetospheric configuration, the
calculation of the spin-down luminosity of force-free oblique rotators
\cite{2006ApJ...648L..51S}, and the evolution of the obliquity angle
\cite{Philippov:2013aha}, all in flat spacetime. Recently, there have
also been some analytic efforts to understand the emission from an
accelerated isolated pulsar in flat spacetime (see
e.g. \cite{Brennan:2013ppa,Gralla:2014yja}).

In addition to assuming that pulsars possess dipole magnetic fields,
the common assumptions and simplifications in these  earlier
studies have been the following:
\begin{enumerate}
\item \label{ass1} The pulsar magnetosphere is well described by 
force-free electrodynamics.

\item \label{ass2} The backreaction of the magnetic field onto the
  interior matter is ignored.

\item \label{ass3} Deviations from sphericity of the stellar surface
 (e.g. due to rapid rotation) are ignored.

\item \label{ass4} Curved spacetime effects are ignored. 
\end{enumerate}

For determining the pulsar spin-down luminosity, assumption \ref{ass1}
is further justified by new results reported in
\cite{Philippov:2013tpa}, where a particle-in-cell simulation of the
global pulsar magnetosphere yields a spin-down luminosity consistent
with earlier force-free studies~\cite{2006ApJ...648L..51S}.  While
assumption \ref{ass2} is likely to break down in a thin layer near the
stellar surface~\cite{Mestel94,Contopoulos:1999ga,Morrison:2004fp},
the calculations reported in ~\cite{Palenzuela:2012my} indicate that
the backreaction of the magnetic field can lead only to small
corrections to the spin-down luminosity. To our knowledge the extent
to which assumption \ref{ass3} is correct has never been studied
before (likely due to the complexity of imposing boundary conditions
on a nonspherical stellar surface). Relaxing assumption \ref{ass4} has
been argued to lead to important effects.  For example,
\cite{Deutsch55} pointed out that general relativity (GR) can amplify
the induced electric field due to the rotation of the star.  It has also
been argued \cite{Beskin90,Musmilov92} that frame dragging induces an
enhanced electric field which contributes to particle acceleration in
the polar cap region.  More recently, the {\it vacuum} Maxwell
equations were solved in the curved spacetime of a {\it slowly}
rotating NS endowed with a dipole magnetic
field~\cite{Petri:2014oza}. It was shown that GR results in an
enhancement in the spin-down luminosity of a pulsar in vacuum up to
$\sim 34\%$ (see also \cite{Rezzolla:2004hy} for an earlier analytic study). 
Furthermore, adopting a GR resistive MHD scheme, it has been reported that the
spin-down luminosity from a GR aligned rotator deviates from the
corresponding flat spacetime value by $\sim
20\%$~\cite{Palenzuela:2012my}.  As noted in ~\cite{Palenzuela:2012my}
this deviation could be due to the adopted resistive MHD scheme, to the
way the flat spacetime spin-down luminosity formula is applied when
dealing with oblate stars or to GR effects. The origin of the
difference could also be due to the inclusion of backreaction
of the magnetic fields onto the thin layer near the NS surface or some
combination of all these factors. Finally, force-free simulations
of rotating and nonrotating, collapsing neutron stars have recently
been performed in \cite{Lehner:2011aa}. All these earlier studies
indicate that GR plays an important role in pulsar magnetospheres.

In this paper we perform the first systematic study of general
relativistic effects in the force-free pulsar magnetospheres of aligned
rotators. We adopt ideal GRMHD for the NS interior, and model its
magnetosphere by adopting general relativistic force-free
electrodynamics. Using the numerical method we presented and tested
in~\cite{Paschalidis:2013gma} we match the ideal MHD dense interior to
the force-free exterior. We evolve the stars until the systems relax
and compute the spin-down luminosity after steady state has been
achieved.  As in most previous studies we neglect the back-reaction of
the magnetic field onto the matter interior. Consequently, all our results
scale with the magnetic field strength (or the dipole magnetic moment).

We split our study into two stages: in the first stage we treat the
``slow'' rotation limit. In particular we generalize the flat
spacetime results for the spin-down luminosity by considering
sequences of Tolman-Oppenheimer-Volkoff (TOV) spherical stars,
endowing these stars with a slow, uniform, rotational velocity, and a
general relativistic dipole magnetic field~\cite{Wasserman83}.
When both $T/|W| \ll 1$ and ${\cal M}/|W| \ll 1$, where $T$, ${\cal
  M}$ and $|W|$ are the kinetic, magnetic and gravitational binding
energies, respectively, the stellar structure remains unaffected
and nearly spherical.
These simulations allow us to isolate and quantify the effects of the
{\it compaction} on the pulsar spin-down luminosity, as frame dragging is not
present. We find that the higher the compaction, the larger the
enhancement in the spin-down luminosity when compared to its flat
spacetime value. The maximum enhancement reaches $\sim 9.3\%$ near
the maximum compaction limit.

In the second stage, we treat the ``rapid'' rotation limit by
constructing self-consistent equilibrium sequences of uniformly
rotating, polytropic NSs using the Cook-Shapiro-Teukolsky (CST)
code~\cite{Cook:1993qj,Cook:1993qr}.  Endowing these stars with a
general relativistic dipole magnetic field (assuming ${\cal M}/|W| \ll
1$, hence leaving the nonspherical structure of the rotating star
unaffected by the magnetic field) and evolving until relaxation, we 
find that rotation (frame dragging) increases the enhancement of 
the spin-down luminosity over the corresponding Minkowski value even 
further. The maximum enhancement for $n=1$ polytropes is $\sim 23\%$. 
Evolving a rapidly rotating $n=0.5$ polytrope we find an even greater 
enhancement of $\sim 35\%$. Using our simulation data, we derive fitting 
formulas for the pulsar spin-down luminosity as a function of the NS compaction,
angular speed, and dipole magnetic moment. We argue that general
relativistic effects are responsible for the observed enhancement and
not the distorted surface of the star by considering the flat
spacetime evolution of a highly oblate, magnetized star. We find that
the spin-down luminosity in this case is practically the same as for a
spherical star. We expect that stiffer equations of state and more
rapidly rotating stars should lead to even larger enhancements in the
pulsar spin-down luminosity.

This paper is organized as follows. In
Sec.~\ref{sec:numerical_setup} we briefly summarize our numerical
method for evolving and matching ideal GRMHD to its force-free
electrodynamics (GRFFE) limit. We also present the diagnostics we adopt to
monitor these systems. A detailed description of our initial data is
presented in Sec.~\ref{sec:ID}. A set of tests and 
results are summarized in Sec.~\ref{sec:results}, where we also
include a resolution study. We conclude in
Section~\ref{sec:conclusions} with a summary  and
a discussion of  astrophysical implications. Throughout we adopt
geometrized units where $G=c=1$.

%%%%%%%%%%%%%%%%%%%%%%%%%%%%
%%% numerical approach
%%%%%%%%%%%%%%%%%%%%%%%%%%%%

\section{Numerical method}
\label{sec:numerical_setup}
We now briefly describe our numerical technique for matching ideal
GRMHD to its force-free limit. A detailed description of this
technique was presented and tested employing a robust suite of tests in
\cite{Paschalidis:2013jsa,Paschalidis:2013gma}.  We also describe the
adopted grid hierarchy of our numerical evolution, as well as the diagnostic
quantities used to extract relevant physical information.

%%%%%%%%%%%%%%%%%%%
%%%% Force-Free %%%
%%%%%%%%%%%%%%%%%%%
\subsection{Ideal MHD-FFE matching}

The force-free approximation can be used when the electromagnetic
energy density dominates over the matter energy density, as in the
magnetospheres of black holes or NSs.  Considering that  FFE is a
limit of ideal MHD, the ideal MHD equations [see
e.g. Eqs.~(50)-(53)~in \cite{Paschalidis:2013gma}] can be used to
evolve both a perfectly conducting, dense fluid and an extremely
low-density plasma, treating the latter in the force-free
approximation.  Exploiting this correspondence, we have developed and
tested a new numerical scheme for matching dense, ideal MHD stellar
interiors to force-free magnetospheric
exteriors~\cite{Paschalidis:2013gma} in GR when the rest-mass density
distribution, velocity and the spacetime metric are known (see also
\cite{Lehner:2011aa} for an alternative matching technique).  In dense,
ideal MHD regions our method simply advects the magnetic field via the
magnetic induction equation (the ``frozen-in'' condition). The
interior electric field and the Poynting vector follow from the ideal
MHD condition [see e.g. Eq.~(A1) in~\cite{Paschalidis:2013gma}]. The
interior fields at the surface of the star are smoothly matched to
their force-free exterior values.  

The force-free dynamical variables we adopt are the magnetic field
$\mathbf{B}$ and the Poynting vector $\mathbf{S}$. In terms of these
variables, the force-free constraints ${}^*F^{\mu\nu} F_{\mu\nu} = 0$
and $F^{\mu\nu}F_{\mu\nu}>0$~\cite{Blandford02}, where $F_{\mu\nu}$ is
the Faraday tensor, become $\mathbf{S}\cdot\mathbf{B}=0$ and
$\mathbf{B}^4-\mathbf{S}^2>0$ \cite{Paschalidis:2013gma}. The
evolution equations for these dynamical variables are written  as
a set of conservation laws precisely in the same form as the ideal MHD
evolution equations \cite{McKinney:2006sc,Paschalidis:2013gma}, so
that the same GRMHD infrastructure can be adopted to solve both the
ideal MHD and the FFE equations.

When treating pulsars with sufficiently weak magnetic fields (${\cal
  M}/|W| \ll 1$) the matter  and velocity profiles and the
spacetime metric can be determined to high approximation by solving the
Einstein equations for a stationary gravitational field in axisymmetry,
coupled to the equation of hydrostatic equilibrium. Since in this work
we assume that the magnetic fields are weak, the background fluid and
metric fields are kept fixed and correspond to stationary,
axisymmetric, uniformly rotating neutron stars. We thus only need to
evolve the electromagnetic fields in these stationary background matter
and gravitational fields.

\subsection{Evolution method}
\label{subsec:numerical_setup}

Our force-free formulation is embedded in the Illinois GRMHD adaptive
mesh refinement code. This code has been extensively tested and
presented in different scenarios involving compact objects and/or
electromagnetic fields
\cite{Duez:2005sf,Etienne:2010ui,Etienne:2011re,PhysRevLett.109.221102,Gold:2013zma}.
The force-free module solves the equations of ideal GRMHD/FFE adopting
high-resolution shock-capturing methods. Here we employ PPM
reconstruction~\cite{Colella84} coupled to the Harten, Lax and van
Leer approximate Riemann solver~\cite{Harten83}.

To enforce the $\nabla\cdot\mathbf{B}=0$ constraint, the magnetic
induction equation is solved using a vector potential formulation (see
~\cite{Etienne:2011re} for details), coupled to the generalized Lorenz
gauge condition we developed in~\cite{PhysRevLett.109.221102}. We choose a
damping parameter $\xi=1.5/\Delta t$, where $\Delta t$ is the time
step of the coarsest refinement level. This condition is designed to
damp and propagate away spurious electromagnetic gauge modes.  The
time integration of all evolution equations is carried out using a
fourth-order Runge-Kutta scheme.

To simultaneously follow the evolution both in the near and the
far zones of the pulsar magnetosphere we adopt a fixed-mesh refinement
grid hierarchy.  The computational grid in all our simulations
consists of seven levels of refinement. The length of each refinement
box is $2.4\,R_e\times 2^{7-k}$, where $R_e$ is the coordinate
equatorial radius of the NS, and $k=1,\cdots,7$ indicates the level
number. Here, the highest-resolution level corresponds to $k=7$. In a
typical simulation the finest level covers the stellar radius by $68$
zones. The outer boundary is located approximately twenty NS
radii beyond the light cylinder. We set the Courant factor 
$\Delta t/\Delta x=0.45$ for $k=6,7$
and $\Delta t/\Delta x=0.45/2^{5-k}$ for $k=1,\cdots, 5$.

%%%%%%%%%%%%%%%%%%%%%
%%%  Diagnostics  %%%
%%%%%%%%%%%%%%%%%%%%%

\subsection{Diagnostics}
\label{subsection:dia_g_L}
We compute the outgoing Poynting luminosity using both the Newman-Penrose
scalar $\phi_2$ and the Poynting vector ${\bf S}$ \cite{Paschalidis:2013jsa},
\begin{equation}
L\equiv\frac{1}{4\,\pi}\,
\lim_{r\rightarrow\infty}
\int r^2\,|\phi_2|^2\,d\Omega=
\lim_{r\rightarrow\infty}
\int r^2\,S^{\hat{r}}\,d\Omega\,.
\label{eq:LEM}
\end{equation}

In all our simulations we compute the Poynting luminosity at several 
radii between the light cylinder and the outer boundary. We find that 
the Poynting luminosity converges with increasing radius, and is already
close to its asymptotic value  for distances between 5 and 10 
light-cylinder radii from the NS center. This value is recorded in Tables ~\ref{tab:TOVID} 
and~\ref{tab:self_rot_ID}.

In stationary and axisymmetric spacetimes one can define the angular
frequency of the magnetic field lines $\Omega_F$, which, within the
light cylinder of the pulsar, must equal the angular frequency of the
star . Hence, we also monitor $\Omega_F$, which is given 
by~\cite{Blandford02}:
\begin{equation}
\Omega_F(r,\theta)=\frac{F_{tr}}{F_{r\phi}}=
\frac{F_{t\theta}}{F_{\theta\phi}}\,.
\label{eq:angular_freq_O}
\end{equation}

%%%%%%%%%%%%%%%%%
%%% Initial data
%%%%%%%%%%%%%%%%%
\section{Initial data}
\label{sec:ID}

To study the effects of GR on the pulsar spin-down luminosity we split
our study into two stages: In the first stage we treat slowly
rotating NSs and in the second we consider rapidly rotating
stars. Next we describe our initial data. 

%%%%%%%%%%%%%%%%%%%%%%%%%%%%
%% slow rotation limit   %%%
%%%%%%%%%%%%%%%%%%%%%%%%%%%%

\subsection{Slowly rotating stars}
\label{subsection:slow_rotation}

The hydrodynamic and metric data we adopt for ``slowly'' rotating NSs
are solutions of the Tolman-Oppenheimer-Volkoff (TOV) equations; see
e.g. \cite{ShaTeu83a}. Assuming that the star is rotating very slowly
($T/|W| \ll 1$), deviations from these spherical solutions are small
and of order ${\cal O}(\Omega^2)$. To model the stellar rotation we
endow these spherical stars with a uniform angular velocity field, in
the same spirit as studies of pulsar magnetospheres in flat spacetime.
Thus, our ``slowly'' rotating pulsar study is the GR generalization of
spherical pulsar magnetospheres in Minkowski spacetime.

For a given angular speed and dipole magnetic moment, our simulations
of magnetospheres depend solely on the compaction $C\equiv M/R$ of the
slowly rotating spherical star, where $M$ is the gravitational (ADM)
mass of the star and $R$ its areal radius. In other words, for a given
$C$ we can use {\it any} equation of state to determine the interior
metric and hydrodynamic fields and obtain the {\it same} spin-down
luminosity. This is expected because a) in GR, the spacetime outside a
spherical star of compaction $C$ is the same, independent of the
interior metric and structure, b) the stars are slowly rotating so
that the light cylinder radius is much larger than the stellar radius,
and its boundary lies in the flat spacetime regime, and c) the
magnetosphere within the light-cylinder corotates with the star,
preserving its dipole magnetic field structure. Using $n=0.5$ and $n=1$
polytropic TOV models, we have confirmed our expectation that the
calculated spin-down luminosity is independent of the interior model
for the NS, and depends only on the value of $C$. Thus,  we adopt
a sequence of (analytic) incompressible TOV stars (see e.g.
\cite{ShaTeu83a}) which allow us to easily sample the allowed range of
stellar compactions for TOV stars.

Table~\ref{tab:TOVID} summarizes the parameters of the sequence of
incompressible stars we consider in this study. Note that, as the
compaction of the star increases toward the maximum value $4/9$($=0.44\bar4$), the
central pressure and metric begin to blow up, and when $C=4/9$ the spacetime
becomes singular~\cite{ShaTeu83a}. 

%%%%%%%%%%%%%%%%%%%
%%%% TOV table  %%%
%%%%%%%%%%%%%%%%%%%
\begin{table}[h]
\caption{\label{tab:TOVID} Properties of TOV stars. We list the
  compaction of the star $C=M/R$, the redshift $Z_p$ of a photon emitted at
  the pole and measured by a static observer at infinity, and
  the spin-down luminosity $L$, in units of the spin-down luminosity of
  a force-free aligned rotator in flat spacetime $L_0 =
  1.02\,\mu^2\Omega^4$.}  \centering
\begin{tabular}{c c c}
\hline\hline
 $C$  & $Z_p$ &$L/L_0$       \\
\hline
    0      &  0     & 1.0    \\
    0.020  &  0.209 & 1.018  \\
    0.080  &  0.091 & 1.048  \\
    0.126  &  0.156 & 1.067  \\
    0.153  &  0.201 & 1.077  \\
    0.211  &  0.315 & 1.085  \\
    0.337  &  0.750 & 1.091  \\
    0.398  &  0.120 & 1.093  \\
\hline\hline
\end{tabular}
\end{table}

Following~\cite{Paschalidis:2013gma}, we endow these stars with a
uniform angular velocity
\begin{align}
v^\phi\equiv \frac{d\phi}{dt}=\Omega=\rm constant,
\label{eq:vphi_TOV}
\end{align}
where $\Omega$ is the angular velocity measured at infinity.
Deviations from strict hydrostatic equilibrium for these
configurations are small since ${\cal M}/|W| \ll 1$ and $T/|W| \ll 1$.
Adopting TOV solutions allows us to isolate and quantify the
effects of the compaction on the pulsar spin-down luminosity, as frame
dragging is not accounted for in the metric.

We choose the angular velocity such that the expected location of the
light cylinder radius $R_{\rm LC}=1/\Omega$ is ten NS radii from the
NS center. As the inner magnetosphere ($r_{\perp} = r\sin\theta \le R_{\rm LC}$) corotates
with the star, the angular velocity of the magnetic field lines
$\Omega_F$ must equal $\Omega$ in this region. As a result the light cylinder in GR
can be computed by finding the locus of points where the speed of the
magnetosphere, as measured, say, by a normal observer, equals the speed
of light.  It is thus the cylindrical surface on which the Lorentz factor $\Gamma =
-n_\alpha u^\alpha$ blows up, where $n^\alpha$ is the unit timelike
vector normal to $t=$const. slices and $u^\alpha$ the four-velocity
corresponding to $v^\phi$. The condition $\Gamma\rightarrow \infty$
gives 
\begin{align}
1= \frac{\gamma_{ij}\,(v^i+\beta^i)\,(v^j+\beta^j)}
{\alpha^2}\,,
\label{eq:RLC}
\end{align}
where $\gamma_{ij}$, $\alpha$ and $\beta^i$ are the 
spatial metric, lapse function and shift vector, respectively, defined 
through the 3+1 decomposition of the spacetime metric
\labeq{}{
\begin{split}
ds^2 = &\ g_{\mu\nu}dx^\mu dx^\nu \\
      = &\ -\alpha^2 dt^2 + \gamma_{ij} (dx^i + \beta^i dt)(dx^j 
+ \beta^j dt)\,.
\end{split}
}
 
As our spacetimes are asymptotically flat, at large distance $R\ll r$,
we have $\alpha \rightarrow 1$, $\beta^i \rightarrow 0$, and $\gamma_{ij}
\rightarrow \delta_{ij}$, and Eq.~\eqref{eq:RLC} reduces to the
well-known flat spacetime result $R_{\rm LC}=1/\Omega$ for the
light-cylinder radius.

Choosing the angular velocity such that
$R_{\rm LC}=10\,R$ makes the computations tractable
and is roughly consistent with our approximation of slow rotation
for the majority of the models we consider here. The ratio of 
the centrifugal to the gravitational acceleration at the equator is
\begin{align}
\frac{a_c}{a_g}\approx \frac{T}{|W|} \approx
\frac{\Omega^2\,R}{M/R^2}=\frac{\Omega^2\,R^3}
{M}=0.01\,C^{-1}\,.
\end{align}
For TOV stars with $C \gtrsim 0.15$, $T/|W| \lesssim 6\%$, hence most
of our stars, and especially the very high compaction ones, are slowly
rotating. While the lower compaction stars are not so
slowly rotating, the rotation rate is sufficiently slow that the light-cylinder
radius is in the asymptotically flat regime, and our slow-rotation study
is simply meant to serve as a generalization to the flat spacetime results. 

%%%%%%%%%%%%%%%%%%%%%%
%%% rotating cases %%%
%%%%%%%%%%%%%%%%%%%%%%
\subsection{Rapidly rotating stars}
\label{subsec:rotating_stars}

The hydrodynamic and metric data for our rapidly rotating compact
stars correspond to equilibrium models of uniformly rotating relativistic NSs 
generated by the Cook-Shapiro-Teukolsky (CST)
code~\cite{Cook:1993qj,Cook:1993qr}.  We adopt a polytropic equation
of state
\labeq{}{ 
P = K \rho_0^{1+1/n}, 
}
where $P$ is the pressure, $\rho_0$ the rest-mass density, and 
$K$ and $n$ are the polytropic constant and index, respectively.

We perform our calculations of rapidly rotating stars in polytropic units,
employing dimensionless quantities as in~\cite{Cook:1993qj,Cook:1993qr}:
\begin{eqnarray}
\bar{M}=&K^{-n/2}\,&M\,,\qquad 
\bar{\Omega}= K^{n/2}\,\Omega\,,\\
\bar{x}^i= K^{-n/2}\,x^i\,, \,\hspace{-0.75cm} & &
\bar{t}= K^{-n/2}\,t\,,\hspace{0.25cm} \bar\mu = K^{-n}\mu\,, 
\end{eqnarray}
where $M$ is the mass, $x^i$ are the spatial coordinates, $t$ is the
time coordinate, and $\mu$ the dipole magnetic moment. Our
calculations scale with $K$, which can be set equal to unity in our
code.

We construct constant $\bar\Omega$ sequences with polytropic index
$n=1$. Table~\ref{tab:self_rot_ID} summarizes the main parameters of
the rotating NS models we consider here. Each constant $\bar\Omega$
sequence consists of models that range from the mass-shedding limit to
the maximum compaction configuration for the given
$\bar\Omega$. Models at the mass-shedding limit are highly oblate,
whereas the deviation from sphericity of models near maximum
compaction is small. We also consider the ``supramassive'' NS limit
for $n=1$~\cite{Cook:1993qr}, and a rapidly rotating $n=0.5$ model. A
supramassive NS is the maximum mass NS configuration for a given
equation of state when allowing for uniform rotation. For $n=1$
such a configuration exceeds the maximum mass of a nonrotating
star by $\lesssim 20\%$.

%%%%%%%%%%%%%%%%%%%%%%
%% magnetic field  %%%
%%%%%%%%%%%%%%%%%%%%%%

\subsection{Electromagnetic fields}
\label{subsection:initial_Bfield}

Flat spacetime studies of pulsar magnetospheres initialize
the exterior NS magnetic field to a dipole magnetic field that
is a solution of the {\it vacuum} Maxwell equations.
Similarly, we assume that our GR stars
possess an exterior dipole magnetic field that at $t=0$ is a 
solution to the vacuum Maxwell
equations in Schwarzschild spacetime. The corresponding toroidal vector potential 
in Schwarzschild coordinates is given by~\cite{Wasserman83}:
\begin{equation}
A_\phi=\frac{3\,\mu\,\sin^2\theta}{4\,M}\,\,\left[1+
\frac{r^2}{2\,M^2}\,\textrm{ln}(1-2\,M/r)+\frac{r}{M}
\right]\,,
\label{eq:A_phi}
\end{equation}
where $\mu$ is the dipole magnetic moment and $M$ is the gravitational
mass of the NS. We use Eq. \eqref{eq:A_phi} to generate the
B-field, both in the interior and exterior of the star. Our chosen
$A_\phi$ is the GR generalization of the A-field used in flat
spacetime studies of pulsar magnetospheres. Strictly speaking, this
generalization is an equilibrium solution only for our spherical stars, because
a pure dipole in the vacuum spacetime outside a rapidly rotating NS is
not given by Eq. \eqref{eq:A_phi}. Given that we know of no analytic
equilibrium solution for a dipole magnetic field in the vacuum spacetime of a rapidly
rotating NS we use Eq. \eqref{eq:A_phi} to  generate the exterior
magnetic field in our rapidly rotating models as well.

We set the initial electric field according to the ideal MHD condition
[see Eq.~(A1) in~\cite{Paschalidis:2013gma}]. To guarantee continuity of the 
electric field across the surface of the star, we set 
the initial velocity $u_i$ to be zero in the exterior except for the perpendicular component to 
the $B$-field, which we choose to fall off as $1/r^2$ from its value at the 
stellar surface. The initial Poynting vector is calculated using
\begin{equation}
S^\mu=-n_\nu\,T^{\mu\nu}_{EM}\,,
\label{eq:Poyn_vector}
\end{equation}
where $T^{\mu\nu}_{EM}$ is the electromagnetic stress-energy tensor
[see e.g. Eq.~(15) in~\cite{Paschalidis:2013gma}].

%%%%%%%%%%%%%%%%%%%%%%%%%%%%%
%%%  Table Rotating cases %%%
%%%%%%%%%%%%%%%%%%%%%%%%%%%%%
\begin{center}
\begin{table}[t]
\caption{\label{tab:self_rot_ID} Properties of uniformly rotating
  stars. We list the angular velocity at infinity $\bar{\Omega}$, the
  period in milliseconds, ${M}\,{\Omega}$, where ${M}$
  is the ADM mass, the compaction $C={M}/{R}$, where ${R}$
  is the equatorial circumferential radius, the ratio of the kinetic to the
  gravitational binding energy $T/|W|$, the polar redshift $Z_p$, the eccentricity
  $e=(1-{R}_p/{R}_e)^{1/2}$, where ${R}_p$ and ${R}_e$ are the proper
  equatorial and polar radii, and the pulsar spin-down luminosity $L$
  in units of its flat spacetime value
  $L_0=1.02\,\mu^2\Omega^2$. 
}
\begin{tabular}{ccccccccc}
\hline\hline
$\bar{\Omega}$ & Period(ms)\footnotemark[1]  & ${M}\,{\Omega}$  & $C$  & $T/|W|$ & $Z_p$  &  $e$     &$L/L_0$\\
\hline
       &      & 0.005  &  0.030   & 0.101  &   0.050   &   0.828  & 1.036\\
       &      & 0.009  &  0.075   & 0.030  &   0.094   &   0.483  & 1.075\\
       &      & 0.013  &  0.132   & 0.012  &   0.173   &   0.317  & 1.096\\
0.10\footnotemark[2]   & 4.05 & 0.015  &  0.161   & 0.008  &   0.222   &   0.263  & 1.110\\
       &      & 0.016  &  0.181   & 0.006  &   0.261   &   0.270  & 1.118\\
       &      & 0.016  &  0.195   & 0.005  &   0.287   &   0.210  & 1.122\\
       &      & 0.015  &  0.244   & 0.002  &   0.406   &   0.136  & 1.129\\
\hline 
       &      &0.0147  &  0.059   & 0.097  &   0.104   &   0.825  & 1.071\\ 
       &      &0.0152  &  0.072   & 0.080  &   0.110   &   0.730  & 1.080\\  
0.15\footnotemark[2]   & 2.70 &0.0211  &  0.135   & 0.028  &   0.189   &   0.463  & 1.115\\
       &      &0.0243  &  0.183   & 0.014  &   0.272   &   0.339  & 1.138\\
       &      &0.0247  &  0.221   & 0.008  &   0.353   &   0.257  & 1.147\\
       &      &0.0231  &  0.243   & 0.004  &   0.404   &   0.207  & 1.149\\      
\hline 
       &      &0.0273  &  0.094   & 0.094  &   0.170   &   0.799  & 1.119\\
       &      &0.0294  &  0.129   & 0.056  &   0.202   &   0.735  & 1.139\\
       &      &0.0318  &  0.160   & 0.036  &   0.245   &   0.520  & 1.152\\
0.20\footnotemark[2]   &2.02  &0.0329  &  0.180   & 0.027  &   0.279   &   0.458  & 1.160\\
       &      &0.0334  &  0.194   & 0.022  &   0.305   &   0.417  & 1.165\\
       &      &0.0333  &  0.221   & 0.014  &   0.360   &   0.342  & 1.170\\
       &      &0.0229  &  0.243   & 0.006  &   0.404   &   0.250  & 1.174\\
\hline 
0.38\footnotemark[2]   & 1.06 &0.070   &  0.183   & 0.081  &   0.403   &   0.777  & 1.230\\
\hline 
0.59\footnotemark[3]   & 0.84 &0.057   &  0.165   & 0.136  &   0.330   &   0.800  & 1.357\\
\hline\hline 
\end{tabular}
\begin{flushleft}
\footnotetext[1]{To assign physical
    units we choose the polytropic constant $K$ such that the
    supramassive limit mass for a given index $n$ equals the supramassive
    limit mass of the Akmal - Pandaripandhe - Ravenhall equation of state
    \cite{Akmal98}, which is $2.46M_{\odot}$ \cite{Morrison:2004fp}.}
\footnotetext[2]{n=1.0}
\footnotetext[3]{n=0.5}
\end{flushleft}
\end{table}
\end{center}

%%%%%%%%%%%%%%%%
%%% Results  %%%
%%%%%%%%%%%%%%%%

\section{Results}
\label{sec:results}
In this section we first present a series of additional 
new tests we performed to check
our code, and then we summarize the results from our numerical
simulations of the aligned rotator models summarized in
Tables.~\ref{tab:TOVID} and \ref{tab:self_rot_ID}. 

\subsection{Tests and calibration}

In \cite{Paschalidis:2013gma} we tested our code and GRMHD-FFE matching
technique using a suite of robust tests both in flat spacetime and 3D
black hole spacetimes. We also reproduced the well-known flat
spacetime, aligned rotator solution. To test the robustness of our code
even further, and to calibrate our curved spacetime solutions, we
performed the following three new tests: i) evolution of a TOV star
endowed with a general relativistic dipole magnetic field and no
rotation, ii) test of corotation of the inner magnetosphere
for a highly compact, slowly rotating TOV star, and (iii) evolution of a highly oblate
star in flat spacetime. 

%%%%%%%%%%%%%%%%%
%%% TOV star %%%%
%%%%%%%%%%%%%%%%%

\subsubsection{TOV star}

\begin{figure}[h]
    \subfigure{\includegraphics[width=0.225\textwidth]{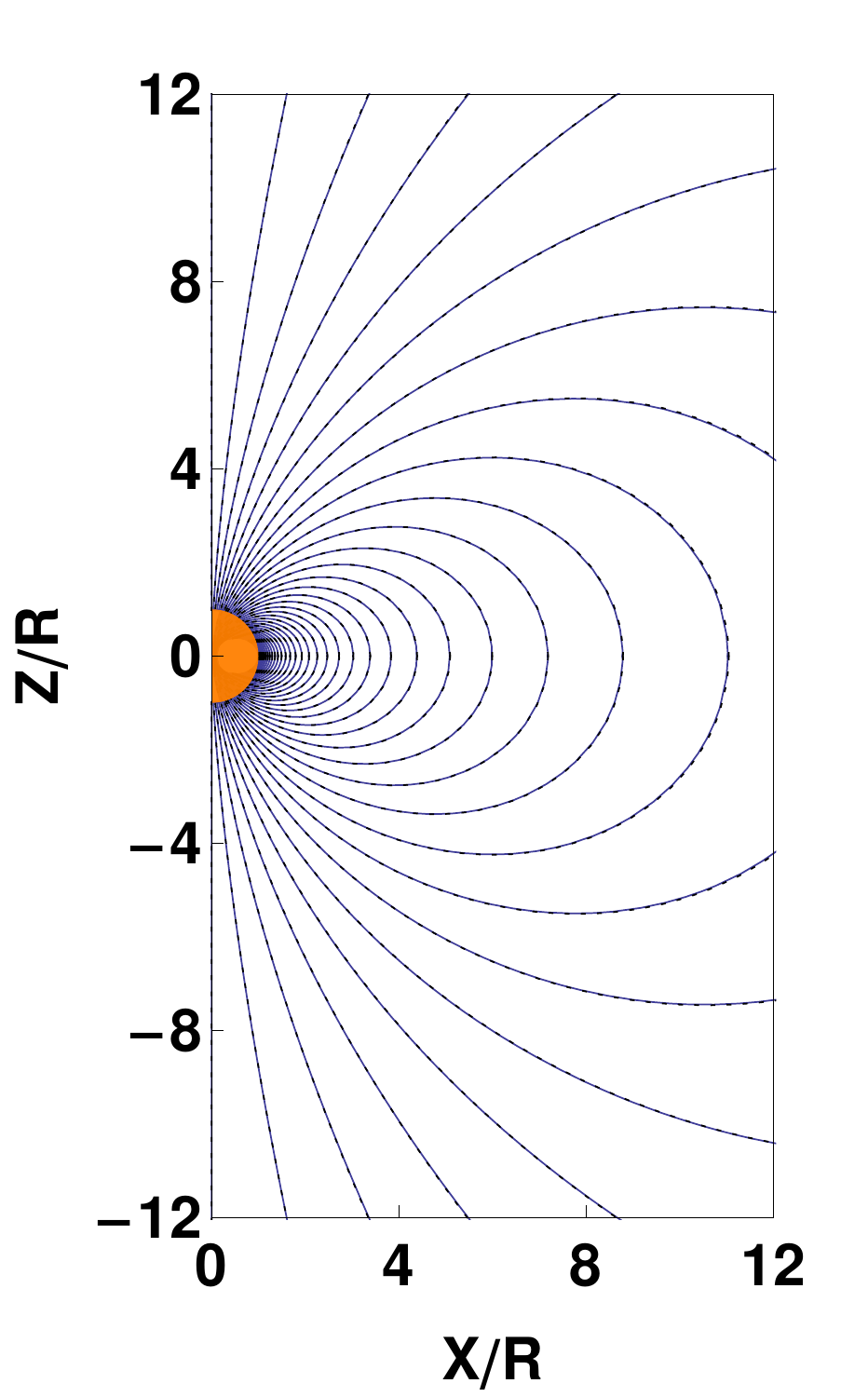}}
    \subfigure{\includegraphics[width=0.23\textwidth]{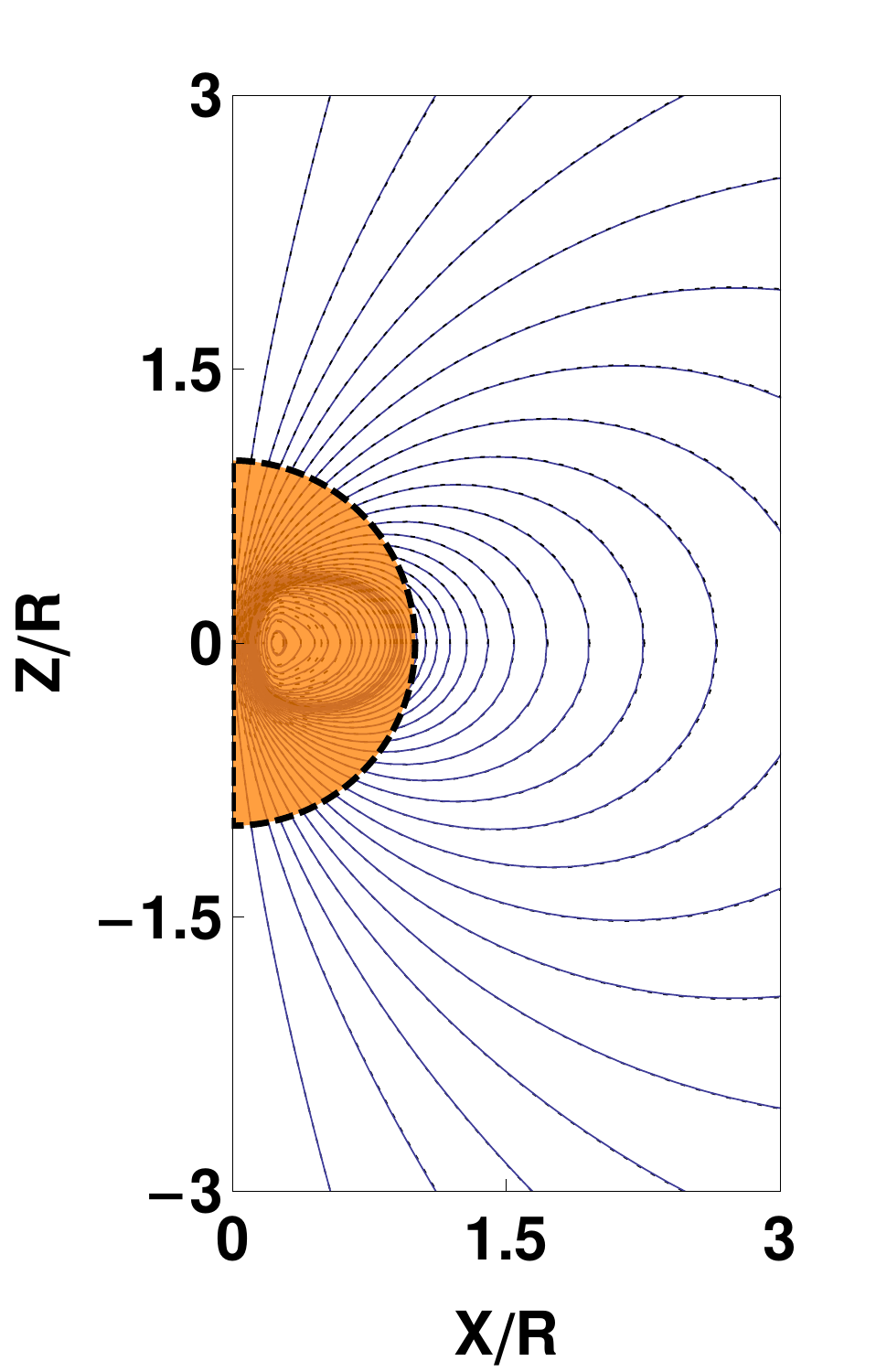}}
   \caption{ \label{fig:TOVstationary} Poloidal magnetic field lines
     for a nonspinning TOV star with compaction $C=0.174$ and endowed
     with a GR dipole magnetic field.  The lines at $t/M=0$ (black
     dashed curves) overlap the field lines at $t/M=442$ (blue solid
     curves). The time $t/M=442$ corresponds to $\sim 2/3$ Alfv\'en
     crossing time.  Left panel: Far zone magnetic field lines. Right
     panel: Near zone magnetic field lines.  }
\end{figure}

A straightforward calculation shows that the solution of the {\it vacuum}
Maxwell equations given in Eq. \eqref{eq:A_phi} satisfies the three
force-free conditions $F_{\mu\nu}J^{\nu} = 0$, ${}^*F^{\mu\nu}
F_{\mu\nu} = 0$ and $F^{\mu\nu}F_{\mu\nu}>0$. As a result
Eq. \eqref{eq:A_phi} is also a solution to the general relativistic
force-free electrodynamic equations.  Consequently, the GRMHD-FFE
evolution of a nonrotating TOV star endowed with a dipole magnetic 
field generated by the vector potential \eqref{eq:A_phi}  must
preserve the initial magnetic field. We have confirmed that this is
indeed the case for a star with compaction $C=0.174$. We evolved 
both the interior and exterior solutions using our code and GRMHD-FFE matching
technique and found that the initial solution is preserved. In
Fig.~\ref{fig:TOVstationary} we plot the poloidal magnetic fields at
$t/M=0$ and $t/M=442$, which corresponds to $2/3$ of an Alfv\'en crossing time
across the computational domain, and demonstrate that the two fields
overlap as expected.

%%%%%%%%%%%%%%%%%%%%%%%%%%%%%%%%%%%%%
%%% FIG: TOV stationary solution %%%%
%%%%%%%%%%%%%%%%%%%%%%%%%%%%%%%%%%%%%

\subsubsection{Corotation of inner magnetosphere} 
\label{corot_TOV}

In all of our evolutions we monitor the angular frequency of the
magnetic field lines $\Omega_F$ computed via
Eq. \ref{eq:angular_freq_O}. To demonstrate that the inner
magnetosphere corotates with the star, even for highly curved
spacetimes, we show here the level of corotation our code achieves for
a TOV star with compaction $C=0.33$ (see Table~\ref{tab:TOVID}).

\begin{figure}[h]
    \subfigure{\includegraphics[width=0.48\textwidth]{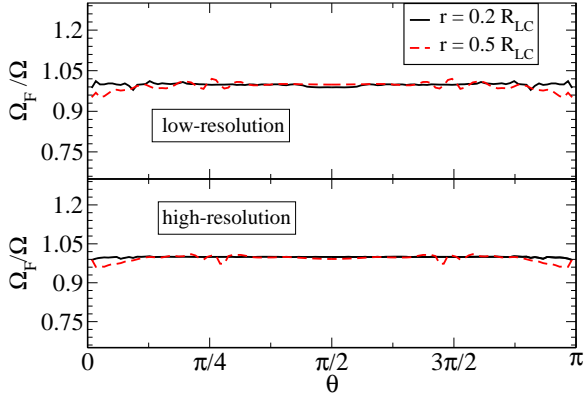}}
   \caption{ \label{fig:angularFrequency} Angular frequency of the
     magnetic field lines $\Omega_F$ normalized by the angular
     velocity of the star ${\Omega}$ as a function of the polar angle
     in the $x-z$ plane at $r =0.2\, R_{LC}$ and $ r=0.5\,R_{LC}$ for
     a TOV star with $C=0.33$. Two different resolutions are shown
     (low and high). As expected, the magnetosphere within the light
     cylinder $R_{LC}$ corotates with star and the higher the
     resolution the higher is the degree of corotation.  }
\end{figure}

Figure~\ref{fig:angularFrequency} plots $\Omega_F$ on the $x-z$ plane
as a function of the polar angle $\theta$ at $r=0.2\, R_{LC}$ and
$r=0.5\,R_{LC}$ at two different resolutions covering the stellar
radius by  $68$ and  $130$ zones, respectively. Here $r$ is the coordinate radius. 
It is clear that the magnetosphere corotates with the star even for 
such high compactions. Notice also that the higher the resolution 
the higher the degree of corotation, as expected from earlier results 
we reported in flat spacetime~\cite{Paschalidis:2013gma}. Similar
levels of corotation are achieved for all models we consider
here.

%%%%%%%%%%%%%%%%%%%%%%%%%%%%%%%
%%% FIG: Angular frequency %%%%
%%%%%%%%%%%%%%%%%%%%%%%%%%%%%%%

%%%%%%%%%%%%%%%%%%
%% oblate star %%%
%%%%%%%%%%%%%%%%%%

\subsubsection{Evolution of an oblate star in flat spacetime} 

To quantify the effects of the shape of the stellar surface on the
structure of the magnetosphere and the spin-down luminosity we evolve
a highly oblate neutron star with eccentricity
$e=(1-{R}_p/{R}_e)^{1/2}=0.799$, where ${R}_p$ (${R}_e$) is the proper
polar (equatorial) radius, that corresponds to the shape and spin of
the equilibrium model near the mass-shedding limit of the sequence
$\bar{\Omega}=0.20$ (see Table \ref{tab:self_rot_ID}). However, we
assume a flat spacetime and endow the star with a flat spacetime
dipole magnetic field.

\begin{figure}[h]
    \subfigure{\includegraphics[width=0.22\textwidth]{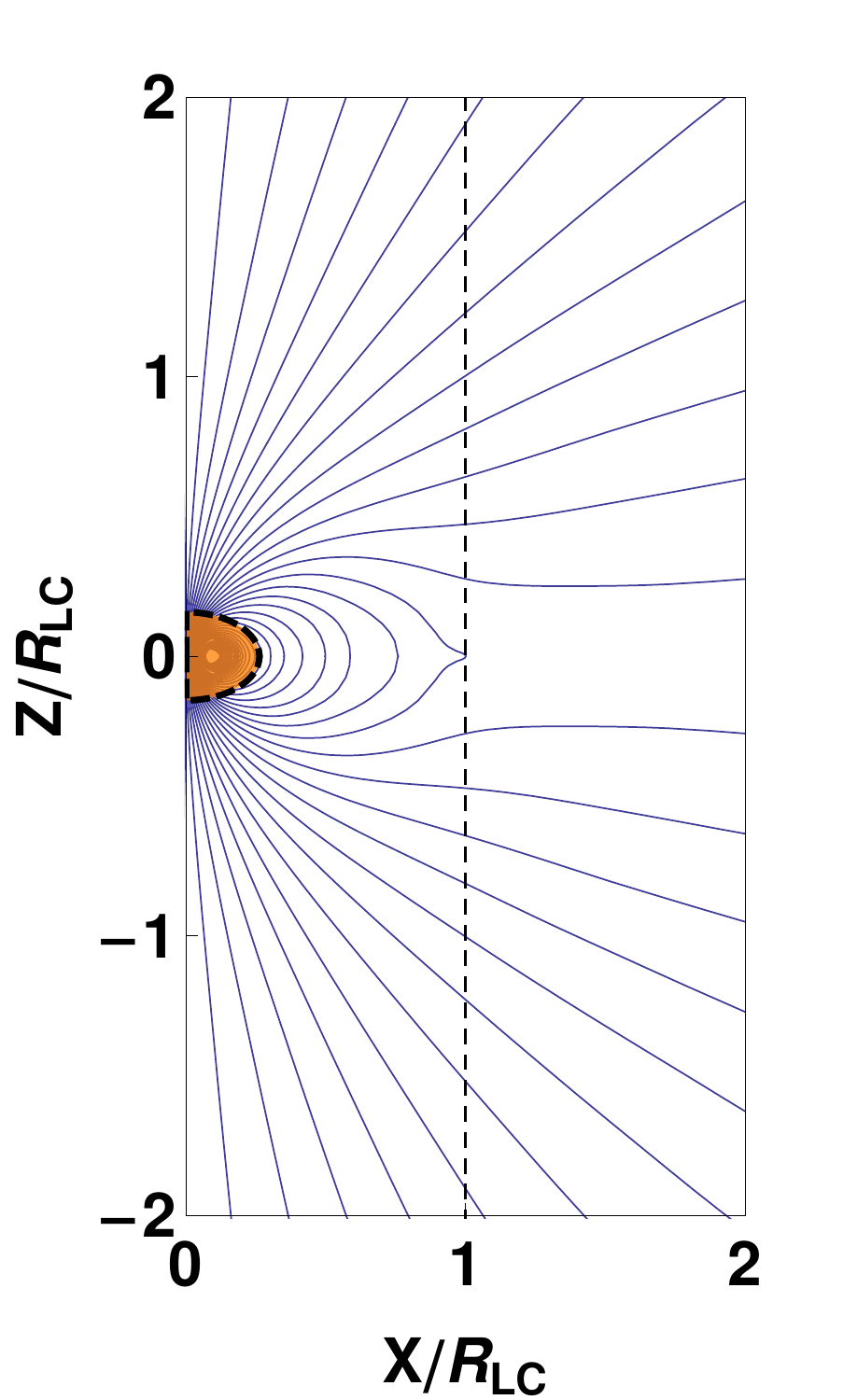}}
    \subfigure{\includegraphics[width=0.244\textwidth]{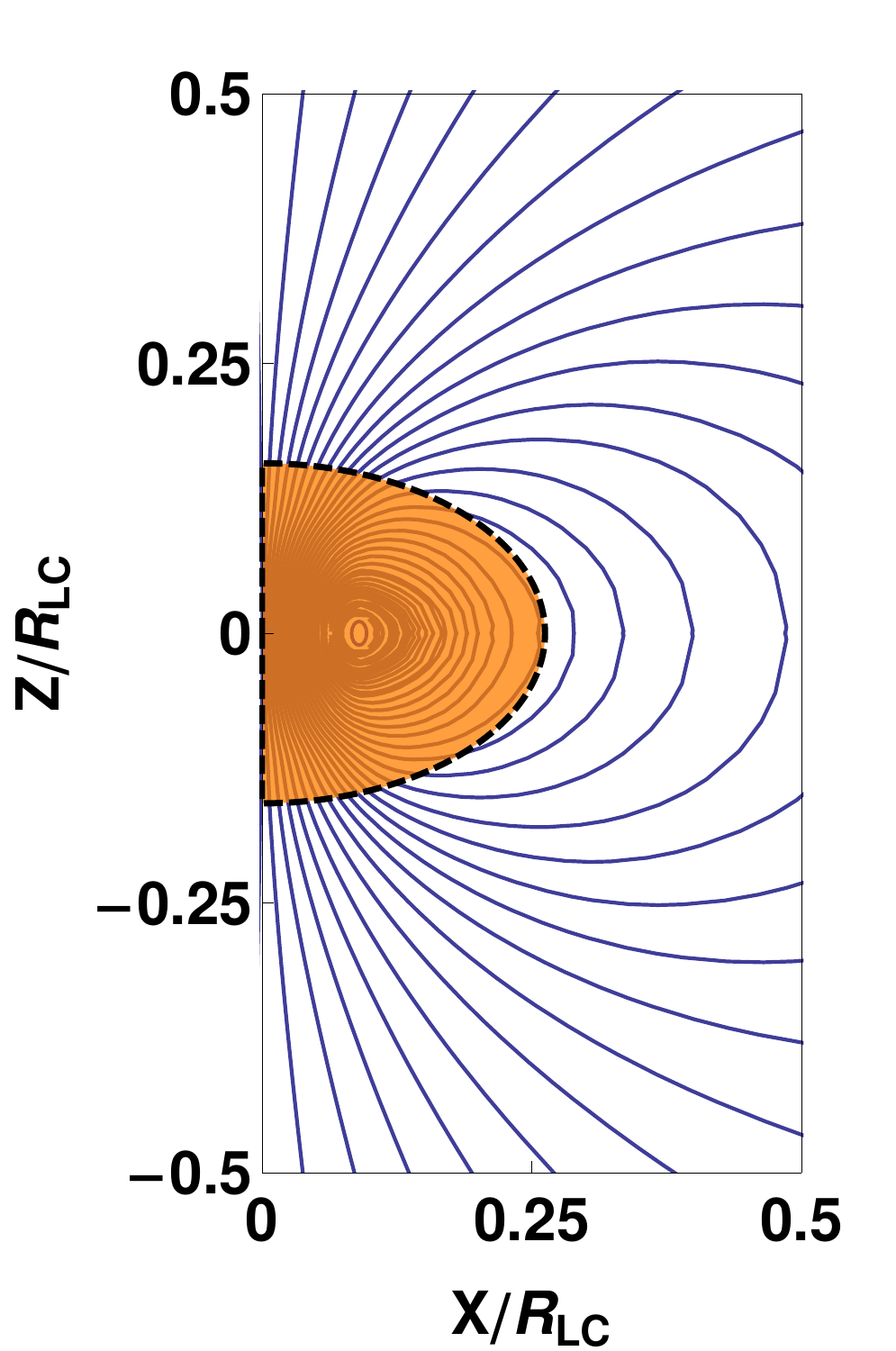}}
   \caption{ \label{fig:poloidalfield_flat} Poloidal magnetic field
     lines on the $ x- z$ plane for a rotating neutron star
     with eccentricity $e=0.799$ and angular velocity $
     \bar{\Omega}=0.20$ (see Table \ref{tab:self_rot_ID}) in flat
     spacetime after the system has relaxed at $ t\approx
     6\pi/{\Omega}$. The shaded areas designate the stellar
     interior.  Left panel: Magnetic field lines in the far zone. The
     dashed vertical line indicates the light cylinder.  Right panel:
     Magnetic field lines in the near zone. }
\end{figure}

The system is evolved until magnetic field relaxation, which occurs
after approximately three rotation periods. The relaxed state has the
same features as the pulsar magnetosphere of a spherical star, i.e.,
an inner closed magnetosphere which corotates with the star, an
equatorial current sheet, and a $Y$ point at the location of the light
cylinder, beyond which the field lines
open. Figure~\ref{fig:poloidalfield_flat} displays the poloidal
magnetic field lines at $ t\approx 6\pi/{\Omega}$ demonstrating the
above features.

%%%%%%%%%%%%%%%%%%%%%%%%%%%%%%%%%%%%%%%%%%%%%%%%%%%%%%%%
%%% FIG: poloidal magnetic field on flat spacetime  %%%%
%%%%%%%%%%%%%%%%%%%%%%%%%%%%%%%%%%%%%%%%%%%%%%%%%%%%%%%%

We find that $\Omega_F$ is equal to $\Omega$ at the same level of
accuracy as reported in Sec.~\ref{corot_TOV}. We find that the pulsar spin-down
luminosity is $L\approx 1.01\,L_0$. Here $L_0$ is the luminosity we
calculate for an independent, high-resolution simulation of a rotating
{\it spherical} star in flat spacetime, 
\labeq{L0}{
  L_0=1.02\,\mu^2\,\Omega^4 \simeq 10^{43} B_{12}^2 R_{10}^6 P_{\rm
    ms}^{-4} \ \ \rm \frac{erg}{sec}, 
} 
where $B_{12} = B/10^{12} \rm G$, $R_{10}=R/10\rm km$ and $P_{\rm ms}
= P/1\rm ms$.  Our Eq. \eqref{L0} is consistent with the value
~$L=(1\pm0.05)\,\mu^2\,\Omega^4$ reported in
\cite{2006ApJ...648L..51S}.

We conclude that for the same $\Omega$ and $\mu$ the outgoing Poynting
luminosity for an aligned rotator in flat spacetime is essentially independent of
the shape of the stellar surface. Notice that this result is not too
surprising, since the magnetic field, both in the interior and the
exterior inner magnetosphere, corotates with the star, preserving the
location of the light cylinder radius beyond which the field lines
open and contribute to the outgoing Poynting flux. Therefore, the
shape of the stellar surface should not matter and the results should
be the same as if the star were a sphere.

This test helps us conclude that deviations from the flat spacetime
spin-down luminosity in GR are likely not due to the distorted surface
of the star, and motivates our looking elsewhere for GR corrections to
the pulsar spin-down luminosity. This is what we do in the following
sections.

%%%%%%%%%%%%%%%%%%%%%
%%%% TOV  Stars  %%%%
%%%%%%%%%%%%%%%%%%%%%
\subsection{Slowly rotating sequence}

We evolved the sequence of TOV stars presented in
Table~\ref{tab:TOVID} until relaxation.  We find that the global
structure of the magnetosphere has the same features as in flat
spacetime (see e.g. Fig. \ref{fig:poloidalfield_flat}) with the main
difference being that the light-cylinder radius is now more consistent
with Eq.~\eqref{eq:RLC}, rather than the flat spacetime value given by
$R_{\rm LC} = 1/\Omega$.

\begin{figure*}[t]
    \subfigure{\includegraphics[width=0.48\textwidth]{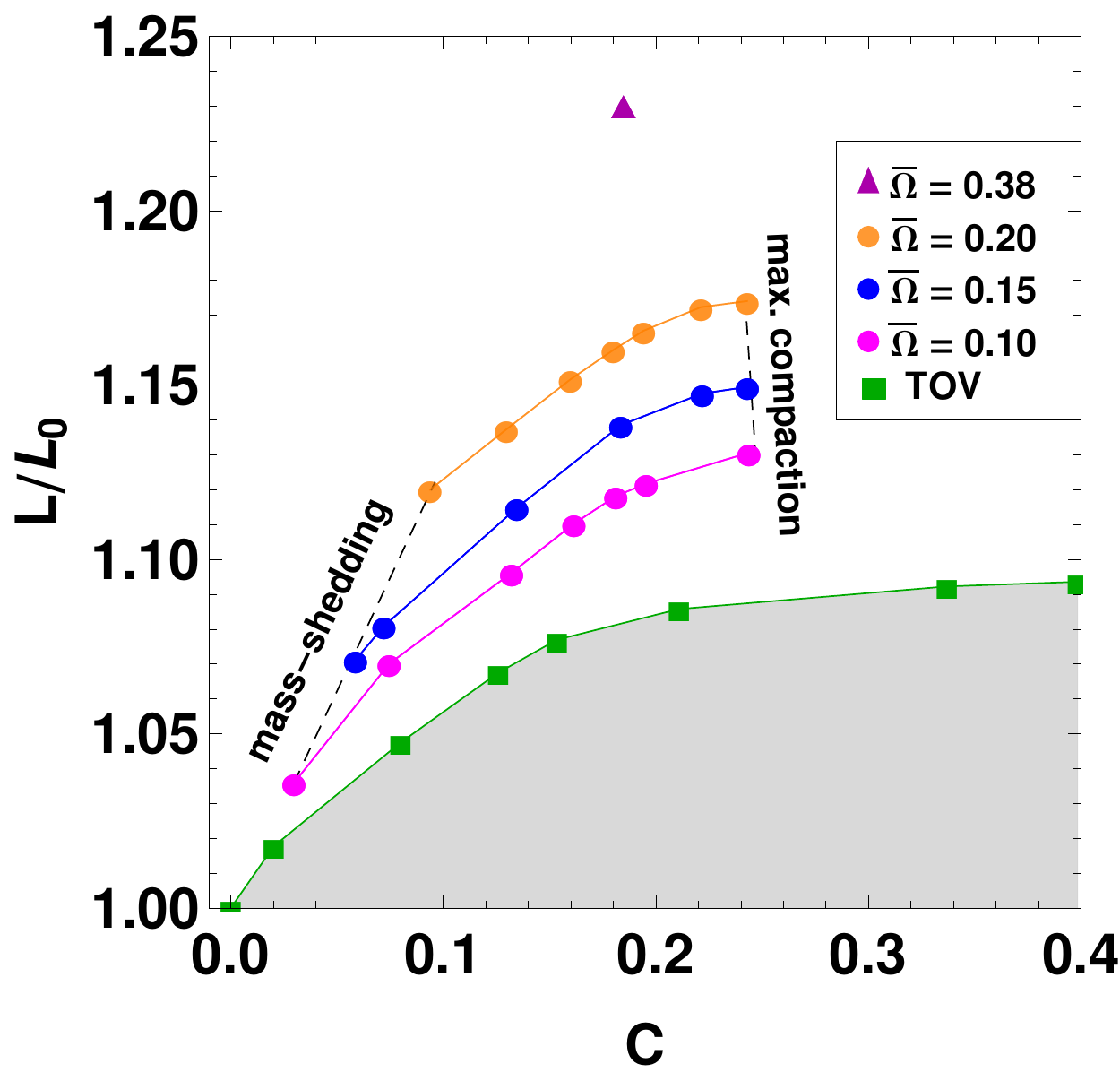}}
    \subfigure{\includegraphics[width=0.46\textwidth]{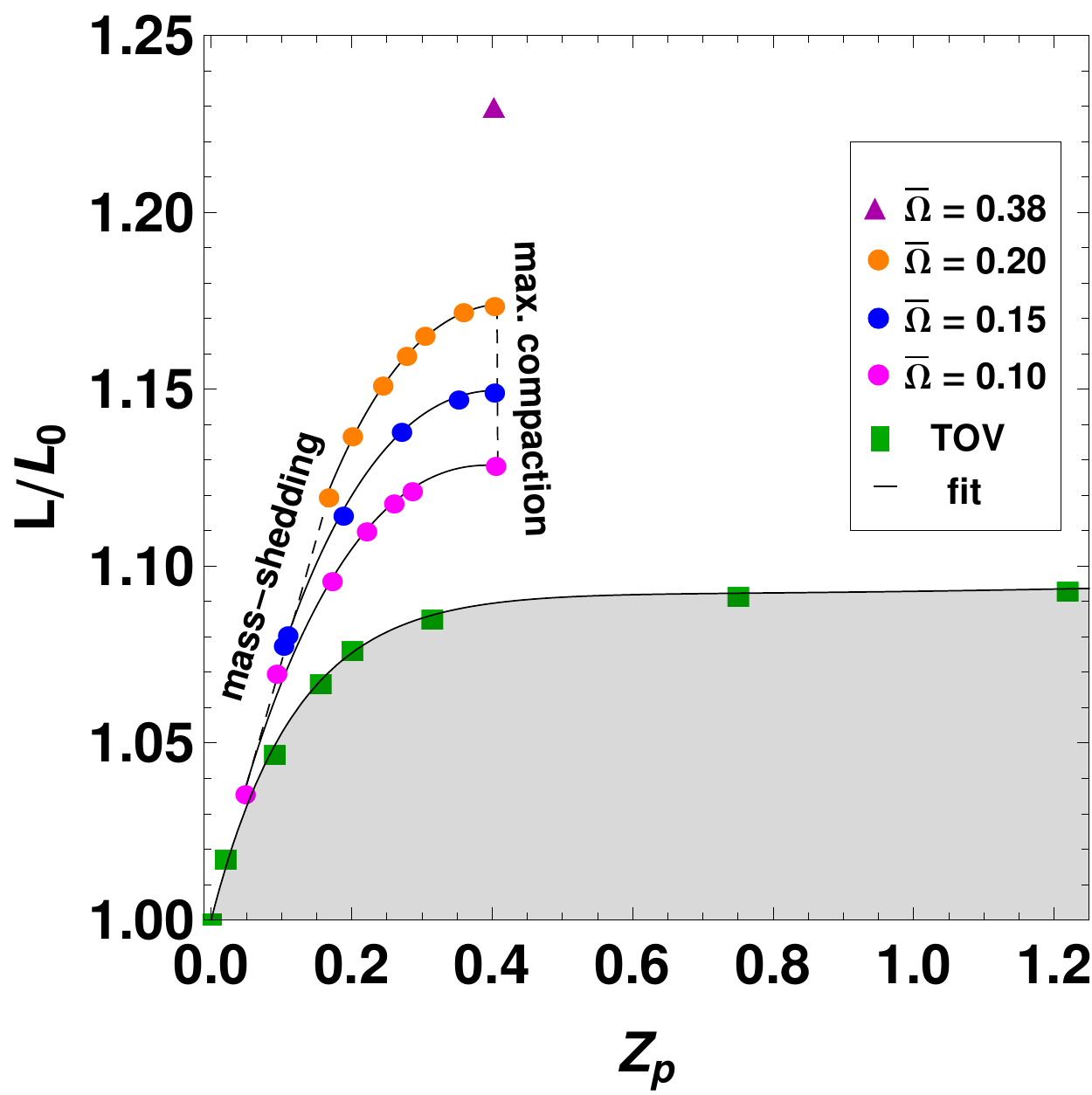}}
    \caption{ \label{fig:luminosity} Pulsar spin-down luminosity $L$
      normalized by $L_0=1.02\mu^2\Omega^4$, our flat-spacetime
      result, for models listed in Tables.~\ref{tab:TOVID}
      and~\ref{tab:self_rot_ID}.  Left panel: $L/L_0$ vs compaction
      $C$. Right panel: $L/L_0$ vs polar redshift $Z_p$, where the
      points are connected by the fitting functions defined in
      Eqs. \eqref{eq:fitTOV}-\eqref{eq:fitRot}.  The parameter space
      for rotating stars is contained between the left dashed line
      (the mass-shedding limit) and the right dashed line (maximum
      compaction). The top point (triangle) corresponds to the
      supramassive neutron star limit for $n=1$. The lower shaded zone
      is the area of the parameter space that cannot be reached,
      unless we assume flat spacetime.}
\end{figure*}

Normalizing the pulsar spin-down luminosity to
$L_0=1.02\mu^2\,\Omega^4$, we find that GR effects enhance the
outgoing Poynting luminosity over $L_0$.  We list the values of
$L/L_0$ for all cases in the last column of Tab.~\ref{tab:TOVID}, and
plot $L/L_0$ vs $C=M/R$, where $M$ is the ADM mass and $R$ the areal
radius of the star, and vs the redshift $Z_p$ from the stellar surface 
as measured by static observer at infinity in
Fig. \ref{fig:luminosity}. As expected, models near the flat spacetime
regime ($C\rightarrow 0$) give rise only to a {\it small} enhancement
of the pulsar spin-down luminosity over $L_0$. However, the
enhancement increases monotonically with $C$, reaching a plateau for
$C\gtrsim 0.35$. For the maximum compaction model we have considered,
the resulting enhancement is approximately $9.3\%$. As our $C\approx
0.4$ model is close to the maximum compaction limit for TOV
stars ($C_{\rm max}=4/9$), we do not expect any further enhancement
beyond $C\approx 0.4$.

%%%%%%%%%%%%%%%%%%%%%%%%%%%%%%%%%%%%%
%%%  FIG: luminosity and omega   %%%%
%%%%%%%%%%%%%%%%%%%%%%%%%%%%%%%%%%%%%

For the TOV sequence we investigated several fitting functions of the
form:
\labeq{}{ 
L=F(C) L_0, 
}
for a given $\Omega$ and $\mu$, and found that the following
fourth-order polynomial provides an excellent fit to the results of
our simulations:
\labeq{eq:fitTOV}{
F(C)=1+ 0.78\,C- 2.16\,C^2-1.77\,C^3+\,0.55\,C^4\,.
}
The right panel in Fig.~\ref{fig:luminosity} shows the above fit, 
where $C$ is implicitly related to $Z_p$ by
\labeq{CofZ}{
Z_p= \bigg(1-\frac{2M}{R}\bigg)^{-1/2}-1= \big(1-2C\big)^{-1/2}-1\,.
}

%\labeq{CofZ}{
%C=\frac{Z_p(Z_p+2)}{(Z_p+1)^2}\,.
%} 

%%%%%%%%%%%%%%%%%%%%%%%%%%%%
%%% rotation sequences   %%%
%%%%%%%%%%%%%%%%%%%%%%%%%%%%
\subsection{Rapidly rotating neutron stars} 

\begin{figure}
  \centering
    \subfigure{\includegraphics[width=0.22\textwidth]{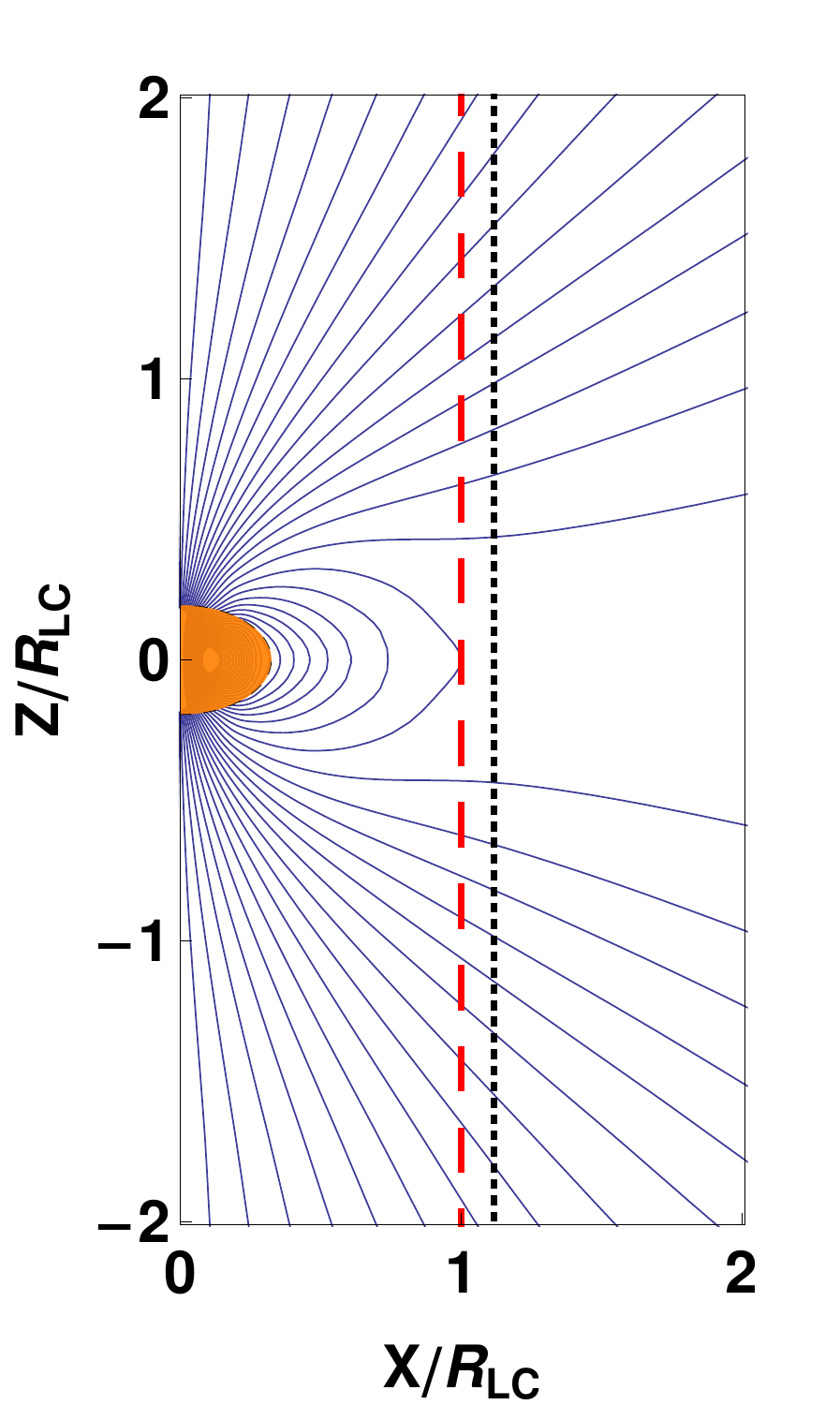}}
    \subfigure{\includegraphics[width=0.245\textwidth]{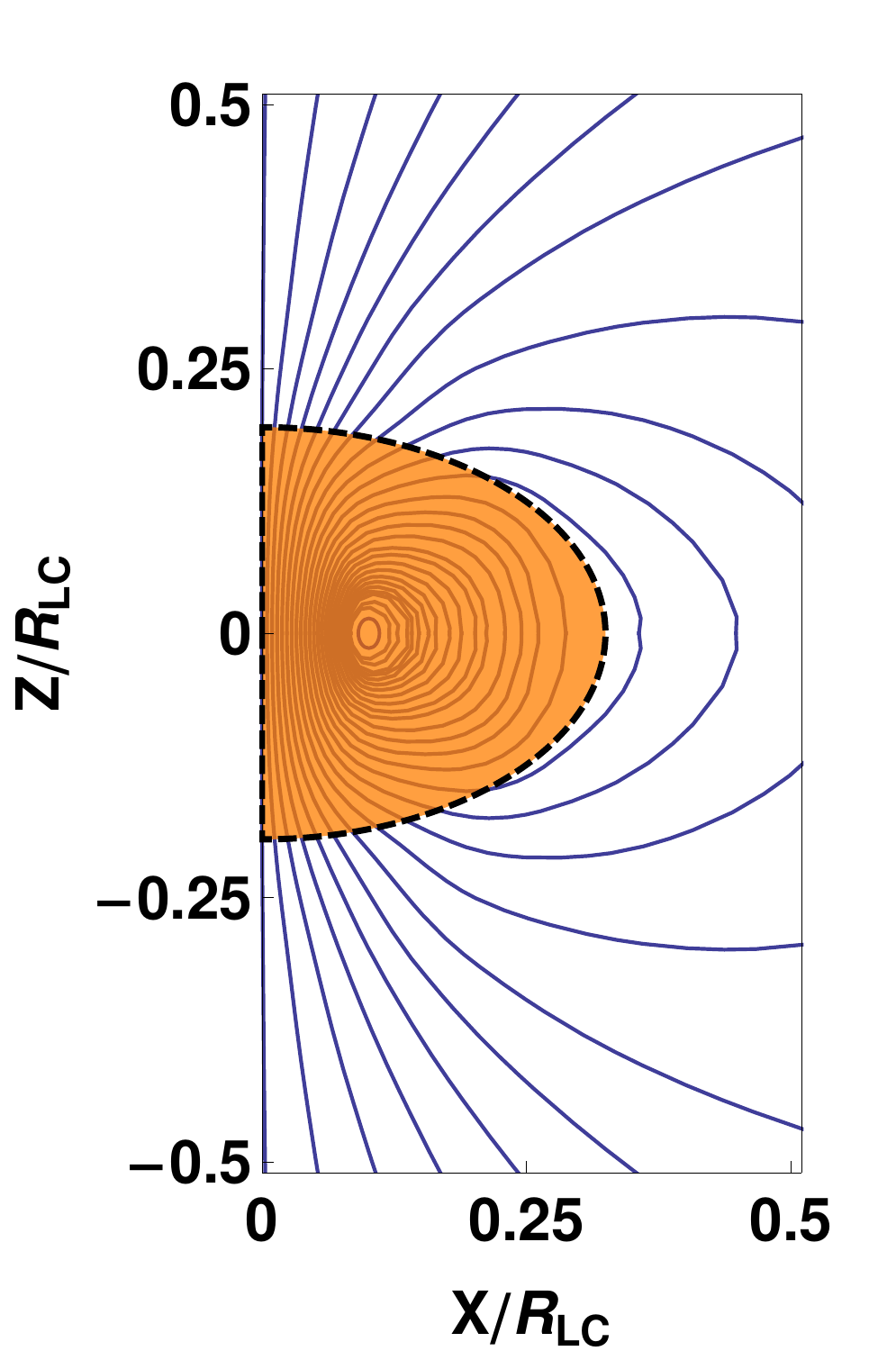}}
    \caption{ \label{fig:Bfield_high} Poloidal magnetic field lines at
      the relaxed state of the rapidly rotating $n=0.5$ NS model (see
      Table~\ref{tab:self_rot_ID}). The shaded area designates the
      stellar interior. Left panel: Far-zone solution. The vertical
      dashed (dotted) line indicates the GR (flat spacetime)
      prediction for the location of the light cylinder. Right panel:
      Near-zone solution. The $Y$ point coincides with the GR
      prediction for the location of the light cylinder.  }
\end{figure}

We evolved the rapidly rotating neutron star models listed in
Table~\ref{tab:self_rot_ID} until relaxation.  The global structure of
the magnetosphere in all cases is similar to the flat spacetime one as
expected.  Figure~\ref{fig:Bfield_high} displays the relaxed poloidal
magnetic field lines on the $ x- z$ plane for the ultrastiff, highly
distorted $n=0.5$ neutron star model (see Table
\ref{tab:self_rot_ID}). The dashed (inner) line indicates the location of the
light cylinder calculated in GR via Eq.~\eqref{eq:RLC}, while the
dotted (outer) line corresponds to the value of $R_{\rm LC}$ in flat
spacetime. The location of the $Y$ point agrees better with the GR
prediction.

Normalizing the pulsar spin-down luminosity $L$ by $L_0$ we find that
accounting for the stellar rotation self-consistently results in a
greater enhancement of $L$ over $L_0$ than the values found in the
previous section. {\it The higher the compaction of the star and the more
rapidly the star rotates, the larger the enhancement}. The maximum
enhancement in $L$ for each of the $\bar{\Omega}=0.10,\, 0.15,\,0.20$
sequences is approximately $13\%$, $15\%$ and $17\%$,
respectively. The maximum enhancement for $n=1$ polytropes we found
here is set by the supramassive limit and is $23\%$.  However,
changing the stiffness of the equation of state we can achieve greater
enhancements. The enhancement in the spin-down luminosity over $L_0$
for the $n=0.5$ model with angular velocity $\bar{\Omega}=0.59$ is
$35\%$. Therefore, {\it we expect stiffer equations of state and more
rapidly rotating neutron stars to lead to even larger enhancements}. We
list the values of $L/L_0$ for all cases in the last column of
Table~\ref{tab:self_rot_ID}, and plot $L/L_0$ vs $C$ and vs $Z_p$
in Fig. \ref{fig:luminosity} for our $n=1$ cases.

%%%%%%%%%%%%%%%%%%%%%%%%%%%%%%%%%%%%%%%%
%%%  FIG: Field lines  rotating NS   %%%
%%%%%%%%%%%%%%%%%%%%%%%%%%%%%%%%%%%%%%%%

For the spin-down luminosity in the $n=1$ rapidly rotating cases we investigated
fitting functions of the form 
\labeq{}{ L = G(Z_p,\bar\Omega) L_0.
} 
We find that the following function
\begin{eqnarray}
G(Z_p,\bar\Omega)=1&+& 0.76\,{\cal C}(Z_p)-0.32\,{\cal C}(Z_p)^2+
7.38\,\bar{\Omega}^2\,{\cal C}(Z_p) \nonumber\\
&-&3.53\,{\cal C}(Z_p)^3-4.75\,\bar{\Omega}^2\,{\cal C}(Z_p)^2\,,
\label{eq:fitRot}
\end{eqnarray}
provides an excellent fit to the simulation data.
Here ${\cal C}(Z_p)$ is an ``effective'' compaction 
function defined as in Eq.~\eqref{CofZ}, 

Equation \eqref{eq:fitRot} has been chosen such that
as $Z_p\rightarrow 0$, $G(Z_p,\bar\Omega) \rightarrow 1$, thereby
recovering the flat spacetime result $L=L_0$. 
Moreover, we considered only even powers in $\bar{\Omega}$
because the resulting spin-down luminosity must not depend
on whether the dipole magnetic moment is aligned or antialigned
with the spin angular momentum of the star.

%%%%%%%%%%%%%%%%%%%%%%%%%%%
%%%  Convergence test   %%%
%%%%%%%%%%%%%%%%%%%%%%%%%%%

\subsection{Resolution study and error bars}

Convergence tests of our GRFFE code and our GRMHD-GRFFE matching
method in black hole spacetimes and in black hole-neutron star
binaries have already been presented in
\cite{Paschalidis:2013jsa,Paschalidis:2013gma}. Here
we perform a convergence test of corotation of the near-zone
magnetosphere using our calculations of the supramassive $n=1$ NS.

We compute the convergence factor $c_F$ 
defined as 
\begin{equation}
c_F=\frac{\|u_{\Delta_1}\|}{\|u_{\Delta_2}\|}\,,
\end{equation}
at two different resolutions $\Delta_1 > \Delta_2$, where we define
$u=1-\Omega_F/\Omega$ at $ r/R_{\rm LC}=0.5$, and
$\|\ \|$ designates the $L_2$ norm.  Using three different resolutions:
the low, medium and high covering the stellar radius by $56$, $68$ and
$85$ zones, respectively, we found that the convergence factor is
$c_F=1.479$, when high and medium resolutions are used, and
$c_F=2.127$, when high and low resolutions are used. These
results imply that our code is $1.8$-order accurate on average, where
the order of convergence $p$ is determined by solving
$c_F=(\Delta_1/\Delta_2)^p$ for $p$.

Studying the spin-down luminosity $L$ as a function of resolution, we
find that $|(L_{\rm high}-L_{\rm low})/L_{\rm high}| \approx 1\%$,
where $L_{\rm high}$ ($L_{\rm low}$) is the spin-down luminosity in
the high (low) resolution. As we have found our results to be
convergent we place an error bar on our high-resolution simulations of
order $2\%$. This implies that the deviations from the flat spacetime
luminosity we reported in the previous section are real and not due to
numerical error.

\begin{figure}[h]
  \subfigure{\includegraphics[width=0.48\textwidth]{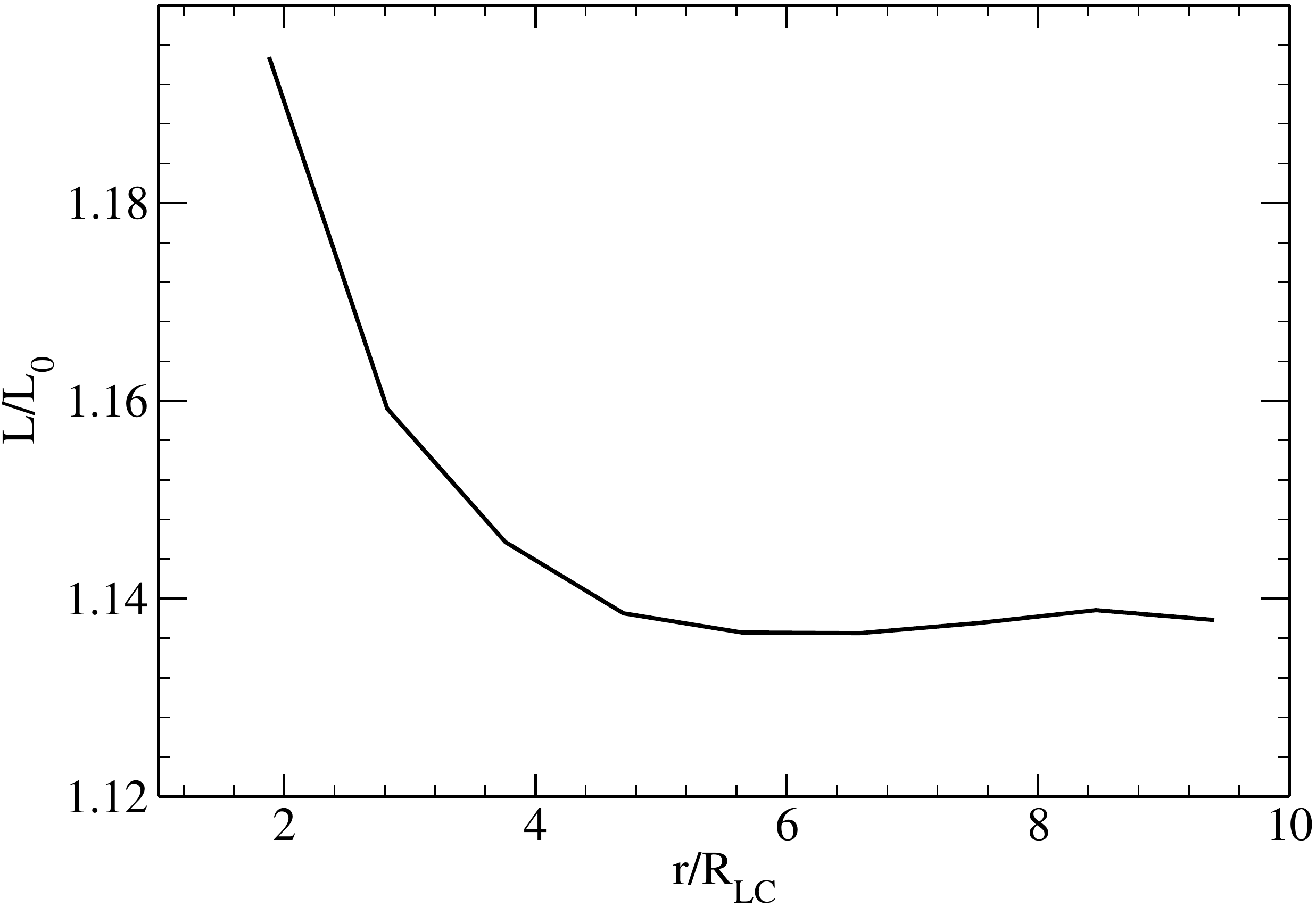}}
  \caption{ \label{fig:luminvsR} Radial dependence  of the Poynting
luminosity $L/L_0$ for an uniformly rotating star with compaction $C=0.183$ and 
angular velocity $\bar{\Omega}=0.15$ (see Table~\ref{tab:self_rot_ID}).
Notice that beyond a radius of $2$ light-cylinder radii and out to $10$ 
light-cylinder radii, the  luminosity drops only by $\sim 4\%$ which
indicates that the dissipation must be small.}
\end{figure}

As pointed out by~\cite{Tchekhovskoy:2012hm} and~\cite{Philippov:2013aha}, 
some force-free and MHD studies of  aligned rotators, have strong numerical 
dissipation of the Poynting luminosity past the light cylinder. Defining 
the dissipation of the Poynting luminosity in GR in the same way as in flat 
spacetime studies (see e.g.~\cite{Tchekhovskoy:2012hm}), is not meaningful 
because of gauge ambiguities that may arise due to calculating the luminosity 
as a surface integral over a sphere  of constant radius in the strong-field 
regime. To show how much dissipation may exist in our simulations,  we plot 
in Fig.~\ref{fig:luminvsR} the radial dependence  of the Poynting
luminosity  for a uniformly  rotating star with compaction $C=0.183$ and 
angular velocity $\bar{\Omega}=0.15$ (see Table~\ref{tab:self_rot_ID}),  where 
the light cylinder is in a regime where the gravitational field is not too 
strong. Past a radius of 
$2$ light-cylinder radii and out to $10$ light-cylinder radii, the  luminosity 
drops only by $\sim 4\%$, asymptoting to the value we quote in 
Table~\ref{tab:self_rot_ID}. It is not entirely clear whether the drop is due 
to numerical dissipation of the Poynting luminosity, to gauge contributions 
in the surface integral, or to a combination of the two. However, if it is due 
to dissipation, the result indicates that the dissipation must be small 
and near our quoted error bars.

%%%%%%%%%%%%%%%%%%%%
%%% Discussion   %%%
%%%%%%%%%%%%%%%%%%%%

\section{Conclusions}
\label{sec:conclusions}

Magnetized neutron stars possess a force-free magnetosphere.  Pulsar
magnetospheres have been studied numerically over the last two
decades. However, all of these early force-free studies were carried
out in flat spacetime. It has been suggested that general relativistic
effects may become important but due mainly to frame
dragging~\cite{Beskin90,Musmilov92}.
 
Using the new method we recently developed for matching general
relativistic ideal MHD to its corresponding force-free limit, we have
performed the first systematic study of aligned rotator force-free
magnetospheres in GR. We constructed equilibrium sequences of
``slowly'' and rapidly rotating NSs. The former are modeled
as incompressible TOV stars covering almost the entire allowed range
of compaction. The latter are constant $\Omega$ sequences of
uniformly, rapidly rotating, GR, polytropic stars generated by the CST
code, and ranging from the mass-shedding limit to the maximum
compaction configuration for a given stellar angular frequency
$\Omega$. We endowed these stars with a weak GR dipole magnetic field
and evolved the fields to relaxation.

Some of our rapidly rotating NS models (those close to the mass-shedding limit) are highly distorted (see Table
\ref{tab:self_rot_ID}). To ensure that any deviations from the flat
spacetime spin-down luminosity are due to strong field effects and not to the
distorted surface of the stars, we evolved a highly oblate star (with
eccentricity $e=0.799$) in flat spacetime, and found that the outgoing
Poynting luminosity is independent of the shape of the stellar surface,
i.e., the resulting spin-down luminosity is approximately the same
as if the star were a sphere spinning with the same angular frequency
and endowed with the same magnetic field. 

Normalizing the spin-down luminosity $L$ to its corresponding Minkowski
value $L_0$, we find that GR effects give rise to a modest
enhancement. As both the compaction and the stellar angular frequency
increase, the enhancement becomes more pronounced. In the ``slow''
rotation limit, where we isolate the effects of compaction, the maximum enhancement is $\sim 9.3\%$,
independent of the equation of state. However, for
rapidly rotating stars, where frame dragging is also important, the maximum enhancement for $n=1$ polytropes is
$\sim 23\%$, and for a rapidly rotating $n=0.5$ polytrope we find an
enhancement of $\sim 35\%$.  We expect stiffer equations of state and
more rapidly neutron rotating stars to lead to even larger
enhancements in the spin-down luminosity. For the cases we studied
here we provided fitting functions for the general relativistic
spin-down luminosity of the form $L=G(Z_p,\Omega)L_0$, where $Z_p$
is the gravitational redshift of light emitted from the NS pole
as measured by a static observer at infinity. 

Our results show that even moderate compaction stars (for which 
frame dragging is accounted for) have larger $L/L_0$ than the highest 
compaction TOV model (for which frame dragging is not accounted for). 
Hence, it is natural to conclude that there is a strong correlation
between frame dragging and the enhancement of the spin-down luminosity
over its value in flat spacetime. This suggestion is consistent with 
the findings of earlier theoretical arguments~\cite{Beskin90,Musmilov92} 
of why GR pulsar magnetospheres are different than those in flat spacetime. 
Further studying the source of the  differences between GR and flat spacetime studies 
of  pulsars, is beyond the scope of this paper, and will be the 
subject of a future work.

Should future gravitational wave observations prove able to constrain the nuclear
equation of state (see
e.g. \cite{Lackey:2011vz,Bauswein:2011tp,Paschalidis:2012ff} and
references therein), our study of pulsar magnetospheres and similar
future studies can constrain the NS magnetic field strength and geometry as
follows: pulsars in detached, mildly relativistic binaries allow for a determination of the
pulsar mass, the angular frequency of rotation, and the spin-down
rate. If the nuclear equation of state is known then the compaction of
the NS follows from the mass-radius relationship. 
As the spin-down luminosity depends on the compaction, the
angular speed and the magnetic field, then from observations
one can constrain the magnetic field, given that all the other 
parameters are in principle measurable or can be inferred. Even 
the obliquity angle between the magnetic dipole moment and the rotation 
axis could possibly be constrained, once oblique rotators are studied in 
GR. This motivates an investigation of such rotators in GR, which we intend 
to perform in a future study.

%%%%%%%%%%%%%%%%%%%%%%
%   Acknowledgments
%%%%%%%%%%%%%%%%%%%%%%
\section*{Acknowledgments}

It is a pleasure to thank Roman Gold for useful discussions. This paper
was supported in part by NSF Grants PHY-0963136 and
PHY-1300903 as well as NASA Grants NNX11AE11G and NNX13AH44G at the
University of Illinois at Urbana-Champaign. VP gratefully acknowledges
support from a Fortner Fellowship at UIUC. This work used the Extreme
Science and Engineering Discovery Environment (XSEDE), which is
supported by NSF grant number OCI-1053575. This research is part of
the Blue Waters sustained-petascale computing project, which is
supported by the National Science Foundation (award number OCI
07-25070) and the state of Illinois. Blue Waters is a joint effort of
the University of Illinois at Urbana-Champaign and its National Center
for Supercomputing Applications.

%%%%%%%%%%%%%%%%%%%%%%%%%%%%%%%%%%%%%%%%                                                                                                   
\bibliographystyle{apsrev}           %%%
\bibliography{refs/references}{}     %%%                                                                                                   

%Merlin.mbs v4.21 2009-07-09.
\begin{thebibliography}{10}%
\makeatletter
\providecommand \@ifxundefined [1]{%
 \ifx #1\undefined \expandafter \@firstoftwo
 \else \expandafter \@secondoftwo
\fi
}%
\providecommand \@ifnum [1]{%
 \ifnum #1\expandafter \@firstoftwo
 \else \expandafter \@secondoftwo
\fi
}%
\providecommand \enquote [1]{``#1''}%
\providecommand \bibnamefont  [1]{#1}%
\providecommand \bibfnamefont [1]{#1}%
\providecommand \citenamefont [1]{#1}%
\providecommand\href[0]{\@sanitize\@href}%
\providecommand\@href[1]{\endgroup\@@startlink{#1}\endgroup\@@href}%
\providecommand\@@href[1]{#1\@@endlink}%
\providecommand \@sanitize [0]{\begingroup\catcode`\&12\catcode`\#12\relax}%
\@ifxundefined \pdfoutput {\@firstoftwo}{%
 \@ifnum{\z@=\pdfoutput}{\@firstoftwo}{\@secondoftwo}%
}{%
 \providecommand\@@startlink[1]{\leavevmode\special{html:<a href="#1">}}%
 \providecommand\@@endlink[0]{\special{html:</a>}}%
}{%
 \providecommand\@@startlink[1]{%
  \leavevmode
  \pdfstartlink
   attr{/Border[0 0 1 ]/H/I/C[0 1 1]}%
   user{/Subtype/Link/A<</Type/Action/S/URI/URI(#1)>>}%
  \relax
 }%
 \providecommand\@@endlink[0]{\pdfendlink}%
}%
\providecommand \url  [0]{\begingroup\@sanitize \@url }%
\providecommand \@url [1]{\endgroup\@href {#1}{\urlprefix}}%
\providecommand \urlprefix [0]{URL }%
\providecommand \Eprint[0]{\href }%
\@ifxundefined \urlstyle {%
  \providecommand \doi [1]{doi:\discretionary{}{}{}#1}%
}{%
  \providecommand \doi [0]{doi:\discretionary{}{}{}\begingroup
  \urlstyle{rm}\Url }%
}%
\providecommand \doibase [0]{http://dx.doi.org/}%
\providecommand \Doi[1]{\href{\doibase#1}}%
\providecommand \bibAnnote [3]{%
  \BibitemShut{#1}%
  \begin{quotation}\noindent
    \textsc{Key:}\ #2\\\textsc{Annotation:}\ #3%
  \end{quotation}%
}%
\providecommand \bibAnnoteFile [2]{%
  \IfFileExists{#2}{\bibAnnote {#1} {#2} {\input{#2}}}{}%
}%
\providecommand \typeout [0]{\immediate \write \m@ne }%
\providecommand \selectlanguage [0]{\@gobble}%
\providecommand \bibinfo [0]{\@secondoftwo}%
\providecommand \bibfield [0]{\@secondoftwo}%
\providecommand \translation [1]{[#1]}%
\providecommand \BibitemOpen[0]{}%
\providecommand \bibitemStop [0]{}%
\providecommand \bibitemNoStop [0]{.\EOS\space}%
\providecommand \EOS [0]{\spacefactor3000\relax}%
\providecommand \BibitemShut [1]{\csname bibitem#1\endcsname}%
%</preamble>
\bibitem{ShaTeu83a}%
  \BibitemOpen
  \bibfield{author}{%
  \bibinfo {author} {\bibfnamefont{S.~L.}\ \bibnamefont{Shapiro}}\ and\
  \bibinfo {author} {\bibfnamefont{S.~A.}\ \bibnamefont{Teukolsky}},\ }%
  \emph{\bibinfo {title} {Black Holes, White Dwarfs, and Neutron Stars}}\
  (\bibinfo {publisher} {John Wiley \& Sons},\ \bibinfo {address} {New York},\
  \bibinfo {year} {1983})%
  \bibAnnoteFile{NoStop}{ShaTeu83a}%
\bibitem{lrr-2008-9}%
  \BibitemOpen
  \bibfield{author}{%
  \bibinfo {author} {\bibfnamefont{D.}~\bibnamefont{Psaltis}},\ }%
  \bibfield{journal}{%
  \bibinfo {journal} {Living Reviews in Relativity}\ }%
  \textbf{\bibinfo {volume} {11}} (\bibinfo {year} {2008}),\ \doi{\bibinfo
  {doi} {10.12942/lrr-2008-9}},\ \url{http://www.livingreviews.org/lrr-2008-9}%
  \bibAnnoteFile{NoStop}{lrr-2008-9}%
\bibitem{0264-9381-27-8-084013}%
  \BibitemOpen
  \bibfield{author}{%
  \bibinfo {author} {\bibfnamefont{G.}~\bibnamefont{Hobbs}} \emph{et~al.},\ }%
  \bibfield{journal}{%
  \bibinfo {journal} {Classical and Quantum Gravity}\ }%
  \textbf{\bibinfo {volume} {27}},\ \bibinfo {pages} {084013} (\bibinfo {year}
  {2010}),\ \url{http://stacks.iop.org/0264-9381/27/i=8/a=084013}%
  \bibAnnoteFile{NoStop}{0264-9381-27-8-084013}%
%%CITATION = ARXIV:1305.4385;%%
\bibitem{2002nsps.conf...44S}%
  \BibitemOpen
  \bibfield{author}{%
  \bibinfo {author} {\bibfnamefont{A.}~\bibnamefont{{Shearer}}}\ and\ \bibinfo
  {author} {\bibfnamefont{A.}~\bibnamefont{{Golden}}},\ }%
  in\ \emph{\bibinfo {booktitle} {Neutron Stars, Pulsars, and Supernova
  Remnants}},\ \bibinfo {editor} {edited by\ \bibinfo {editor}
  {\bibfnamefont{W.}~\bibnamefont{{Becker}}}, \bibinfo {editor}
  {\bibfnamefont{H.}~\bibnamefont{{Lesch}}},\ and\ \bibinfo {editor}
  {\bibfnamefont{J.}~\bibnamefont{{Tr{\"u}mper}}}}\ (\bibinfo {year} {2002})\
  p.~\bibinfo {pages} {44},\
  \Eprint{http://arxiv.org/abs/astro-ph/0208579}{astro-ph/0208579}%
  \bibAnnoteFile{NoStop}{2002nsps.conf...44S}%
\bibitem{Atwood:2009ez}%
  \BibitemOpen
  \bibfield{author}{%
  \bibinfo {author} {\bibfnamefont{W.}~\bibnamefont{Atwood}} \emph{et~al.}
  (\bibinfo {collaboration} {LAT Collaboration}),\ }%
  \bibfield{journal}{%
  \Doi{10.1088/0004-637X/697/2/1071}{\bibinfo {journal} {Astrophys.J.}}\ }%
  \textbf{\bibinfo {volume} {697}},\ \bibinfo {pages} {1071} (\bibinfo {year}
  {2009}),\ \Eprint{http://arxiv.org/abs/0902.1089}{arXiv:0902.1089
  [astro-ph.IM]}%
  \bibAnnoteFile{NoStop}{Atwood:2009ez}%
%%CITATION = ARXIV:0902.1089;%%
\bibitem{2005AJ....129.1993M}%
  \BibitemOpen
  \bibfield{author}{%
  \bibinfo {author} {\bibfnamefont{R.~N.}\ \bibnamefont{{Manchester}}},
  \bibinfo {author} {\bibfnamefont{G.~B.}\ \bibnamefont{{Hobbs}}}, \bibinfo
  {author} {\bibfnamefont{A.}~\bibnamefont{{Teoh}}},\ and\ \bibinfo {author}
  {\bibfnamefont{M.}~\bibnamefont{{Hobbs}}},\ }%
  \bibfield{journal}{%
  \Doi{10.1086/428488}{\bibinfo {journal} {Astron. J.}}\ }%
  \textbf{\bibinfo {volume} {129}},\ \bibinfo {pages} {1993} (\bibinfo {month}
  {Apr.}\ \bibinfo {year} {2005}),\
  \Eprint{http://arxiv.org/abs/astro-ph/0412641}{astro-ph/0412641}%
  \bibAnnoteFile{NoStop}{2005AJ....129.1993M}%
\bibitem{FERMILATGPULSARS}%
  \BibitemOpen
  \bibinfo {howpublished}
  {\url{https://confluence.slac.stanford.edu/display/GLAMCOG/Public+List+of+LA%
T-Detected+Gamma-Ray+Pulsars}}%
  \bibAnnoteFile{NoStop}{FERMILATGPULSARS}%
\bibitem{TheFermi-LAT:2013ssa}%
  \BibitemOpen
  \bibfield{author}{%
  \bibinfo {author} {\bibfnamefont{A.}~\bibnamefont{Abdo}} \emph{et~al.}
  (\bibinfo {collaboration} {The Fermi-LAT collaboration}),\ }%
  \bibfield{journal}{%
  \Doi{10.1088/0067-0049/208/2/17}{\bibinfo {journal} {Astrophys.J.Suppl.}}\ }%
  \textbf{\bibinfo {volume} {208}},\ \bibinfo {pages} {17} (\bibinfo {year}
  {2013}),\ \Eprint{http://arxiv.org/abs/1305.4385}{arXiv:1305.4385
  [astro-ph.HE]}%
  \bibAnnoteFile{NoStop}{TheFermi-LAT:2013ssa}%
%%CITATION = ARXIV:1305.4385;%%
\bibitem{2005ApJ...627..397Z}%
  \BibitemOpen
  \bibfield{author}{%
  \bibinfo {author} {\bibfnamefont{S.}~\bibnamefont{{Zane}}}, \bibinfo {author}
  {\bibfnamefont{M.}~\bibnamefont{{Cropper}}}, \bibinfo {author}
  {\bibfnamefont{R.}~\bibnamefont{{Turolla}}}, \bibinfo {author}
  {\bibfnamefont{L.}~\bibnamefont{{Zampieri}}}, \bibinfo {author}
  {\bibfnamefont{M.}~\bibnamefont{{Chieregato}}}, \bibinfo {author}
  {\bibfnamefont{J.~J.}\ \bibnamefont{{Drake}}},\ and\ \bibinfo {author}
  {\bibfnamefont{A.}~\bibnamefont{{Treves}}},\ }%
  \bibfield{journal}{%
  \Doi{10.1086/430138}{\bibinfo {journal} {\apj}}\ }%
  \textbf{\bibinfo {volume} {627}},\ \bibinfo {pages} {397} (\bibinfo {month}
  {Jul.}\ \bibinfo {year} {2005}),\
  \Eprint{http://arxiv.org/abs/astro-ph/0503239}{astro-ph/0503239}%
  \bibAnnoteFile{NoStop}{2005ApJ...627..397Z}%
\bibitem{2006Sci...311.1901H}%
  \BibitemOpen
  \bibfield{author}{%
  \bibinfo {author} {\bibfnamefont{J.~W.~T.}\ \bibnamefont{{Hessels}}},
  \bibinfo {author} {\bibfnamefont{S.~M.}\ \bibnamefont{{Ransom}}}, \bibinfo
  {author} {\bibfnamefont{I.~H.}\ \bibnamefont{{Stairs}}}, \bibinfo {author}
  {\bibfnamefont{P.~C.~C.}\ \bibnamefont{{Freire}}}, \bibinfo {author}
  {\bibfnamefont{V.~M.}\ \bibnamefont{{Kaspi}}},\ and\ \bibinfo {author}
  {\bibfnamefont{F.}~\bibnamefont{{Camilo}}},\ }%
  \bibfield{journal}{%
  \Doi{10.1126/science.1123430}{\bibinfo {journal} {Science}}\ }%
  \textbf{\bibinfo {volume} {311}},\ \bibinfo {pages} {1901} (\bibinfo {month}
  {Mar.}\ \bibinfo {year} {2006}),\
  \Eprint{http://arxiv.org/abs/astro-ph/0601337}{astro-ph/0601337}%
  \bibAnnoteFile{NoStop}{2006Sci...311.1901H}%
\bibitem{Goldreich:1969sb}%
  \BibitemOpen
  \bibfield{author}{%
  \bibinfo {author} {\bibfnamefont{P.}~\bibnamefont{Goldreich}}\ and\ \bibinfo
  {author} {\bibfnamefont{W.~H.}\ \bibnamefont{Julian}},\ }%
  \bibfield{journal}{%
  \bibinfo {journal} {Astrophys. J.}\ }%
  \textbf{\bibinfo {volume} {157}},\ \bibinfo {pages} {869} (\bibinfo {year}
  {1969})%
  \bibAnnoteFile{NoStop}{Goldreich:1969sb}%
%%CITATION = ASJOA,157,869;%%
\bibitem{Michel73}%
  \BibitemOpen
  \bibfield{author}{%
  \bibinfo {author} {\bibfnamefont{F.~C.}\ \bibnamefont{Michel}},\ }%
  \bibfield{journal}{%
  \bibinfo {journal} {Astrophys. J. Letters}\ }%
  \textbf{\bibinfo {volume} {180}},\ \bibinfo {pages} {L133} (\bibinfo {year}
  {1973})%
  \bibAnnoteFile{NoStop}{Michel73}%
\bibitem{Scharleman73}%
  \BibitemOpen
  \bibfield{author}{%
  \bibinfo {author} {\bibfnamefont{E.T.}~\bibnamefont{Scharlemann}}\ and\ \bibinfo
  {author} {\bibfnamefont{R.V.}~\bibnamefont{Wagoner}},\ }%
  \bibfield{journal}{%
  \bibinfo {journal} {Astrophys. J.}\ }%
  \textbf{\bibinfo {volume} {182}},\ \bibinfo {pages} {951} (\bibinfo {year}
  {1973})%
  \bibAnnoteFile{NoStop}{Scharleman73}%
\bibitem{Contopoulos:99}%
  \BibitemOpen
  \bibfield{author}{%
  \bibinfo {author} {\bibfnamefont{I.}~\bibnamefont{Contopoulos}}, \bibinfo
  {author} {\bibfnamefont{D.}~\bibnamefont{Kazanas}},\ and\ \bibinfo {author}
  {\bibfnamefont{C.}~\bibnamefont{Fendt}},\ }%
  \bibfield{journal}{%
  \bibinfo {journal} {Astrophys.J.} \textbf{\bibinfo {volume} {511}},\ \bibinfo {pages} {351}}%
   (\bibinfo {year} {1999})%
  \bibAnnoteFile{NoStop}{Contopoulos:99}%
%%CITATION = INSPIRE-360303;%%
\bibitem{2002astro.ph.11141K}%
  \BibitemOpen
  \bibfield{author}{%
  \bibinfo {author} {\bibfnamefont{S.~S.}\ \bibnamefont{{Komissarov}}},\ }%
  \bibfield{journal}{%
  \bibinfo {journal} {ArXiv Astrophysics e-prints}}%
   (\bibinfo {month} {Nov.}\ \bibinfo {year} {2002}),\
  \Eprint{http://arxiv.org/abs/arXiv:astro-ph/0211141}{arXiv:astro-ph/0211141}%
  \bibAnnoteFile{NoStop}{2002astro.ph.11141K}%
\bibitem{McKinney:2006sd}%
  \BibitemOpen
  \bibfield{author}{%
  \bibinfo {author} {\bibfnamefont{J.~C.}\ \bibnamefont{McKinney}},\ }%
  \bibfield{journal}{%
  \Doi{10.1111/j.1745-3933.2006.00150.x}{\bibinfo {journal}
  {Mon.Not.Roy.Astron.Soc.Lett.}}\ }%
  \textbf{\bibinfo {volume} {368}},\ \bibinfo {pages} {L30} (\bibinfo {year}
  {2006}),\
  \Eprint{http://arxiv.org/abs/astro-ph/0601411}{arXiv:astro-ph/0601411
  [astro-ph]}%
  \bibAnnoteFile{NoStop}{McKinney:2006sd}%
%%CITATION = ASTRO-PH/0601411;%%
\bibitem{2006ApJ...648L..51S}%
  \BibitemOpen
  \bibfield{author}{%
  \bibinfo {author} {\bibfnamefont{A.}~\bibnamefont{{Spitkovsky}}},\ }%
  \bibfield{journal}{%
  \Doi{10.1086/507518}{\bibinfo {journal} {\apjl}}\ }%
  \textbf{\bibinfo {volume} {648}},\ \bibinfo {pages} {L51} (\bibinfo {month}
  {Sep.}\ \bibinfo {year} {2006}),\
  \Eprint{http://arxiv.org/abs/arXiv:astro-ph/0603147}{arXiv:astro-ph/0603147}%
  \bibAnnoteFile{NoStop}{2006ApJ...648L..51S}%
\bibitem{Gruzinov:2008gb}%
  \BibitemOpen
  \bibfield{author}{%
  \bibinfo {author} {\bibfnamefont{A.}~\bibnamefont{Gruzinov}},\ }%
  \bibfield{journal}{%
  \Doi{10.1088/1475-7516/2008/11/002}{\bibinfo {journal} {JCAP}}\ }%
  \textbf{\bibinfo {volume} {0811}},\ \bibinfo {pages} {002} (\bibinfo {year}
  {2008}),\ \Eprint{http://arxiv.org/abs/0804.4176}{arXiv:0804.4176
  [astro-ph]}%
  \bibAnnoteFile{NoStop}{Gruzinov:2008gb}%
%%CITATION = ARXIV:0804.4176;%%
\bibitem{Gruzinov:2013pva}%
  \BibitemOpen
  \bibfield{author}{%
  \bibinfo {author} {\bibfnamefont{A.}~\bibnamefont{Gruzinov}}}%
   (\bibinfo {year} {2013}),\
  \Eprint{http://arxiv.org/abs/1303.4094}{arXiv:1303.4094 [astro-ph.HE]}%
  \bibAnnoteFile{NoStop}{Gruzinov:2013pva}%
%%CITATION = ARXIV:1303.4094;%%
\bibitem{Tchekhovskoy:2012hm}%
  \BibitemOpen
  \bibfield{author}{%
  \bibinfo {author} {\bibfnamefont{A.}~\bibnamefont{Tchekhovskoy}}\ and\
  \bibinfo {author} {\bibfnamefont{A.}~\bibnamefont{Spitkovsky}}}%
   (\bibinfo {year} {2012}),\
  \Eprint{http://arxiv.org/abs/1211.2803}{arXiv:1211.2803 [astro-ph.HE]}%
  \bibAnnoteFile{NoStop}{Tchekhovskoy:2012hm}%
%%CITATION = ARXIV:1211.2803;%%
\bibitem{Uzdensky:2012tf}%
  \BibitemOpen
  \bibfield{author}{%
  \bibinfo {author} {\bibfnamefont{D.~A.}\ \bibnamefont{Uzdensky}}\ and\
  \bibinfo {author} {\bibfnamefont{A.}~\bibnamefont{Spitkovsky}},\ }%
  \bibfield{journal}{%
  \Doi{10.1088/0004-637X/780/1/3}{\bibinfo {journal} {Astrophys.J.}}\ }%
  \textbf{\bibinfo {volume} {780}},\ \bibinfo {pages} {3} (\bibinfo {year}
  {2014}),\ \Eprint{http://arxiv.org/abs/1210.3346}{arXiv:1210.3346
  [astro-ph.HE]}%
  \bibAnnoteFile{NoStop}{Uzdensky:2012tf}%
%%CITATION = ARXIV:1210.3346;%%
\bibitem{Philippov:2013aha}%
  \BibitemOpen
  \bibfield{author}{%
  \bibinfo {author} {\bibfnamefont{A.}~\bibnamefont{Philippov}}, \bibinfo
  {author} {\bibfnamefont{A.}~\bibnamefont{Tchekhovskoy}},\ and\ \bibinfo
  {author} {\bibfnamefont{J.~G.}\ \bibnamefont{Li}}}%
   (\bibinfo {year} {2013}),\
  \Eprint{http://arxiv.org/abs/1311.1513}{arXiv:1311.1513 [astro-ph.HE]}%
  \bibAnnoteFile{NoStop}{Philippov:2013aha}%
%%CITATION = ARXIV:1311.1513;%%
\bibitem{Brennan:2013ppa}%
  \BibitemOpen
  \bibfield{author}{%
  \bibinfo {author} {\bibfnamefont{T.~D.}\ \bibnamefont{Brennan}}\ and\
  \bibinfo {author} {\bibfnamefont{S.~E.}\ \bibnamefont{Gralla}}}%
   (\bibinfo {year} {2013}),\
  \Eprint{http://arxiv.org/abs/1311.0752}{arXiv:1311.0752 [astro-ph.HE]}%
  \bibAnnoteFile{NoStop}{Brennan:2013ppa}%
%%CITATION = ARXIV:1311.0752;%%
\bibitem{Gralla:2014yja}%
  \BibitemOpen
  \bibfield{author}{%
  \bibinfo {author} {\bibfnamefont{S.~E.}\ \bibnamefont{Gralla}}\ and\ \bibinfo
  {author} {\bibfnamefont{T.}~\bibnamefont{Jacobson}}}%
   (\bibinfo {year} {2014}),\
  \Eprint{http://arxiv.org/abs/1401.6159}{arXiv:1401.6159 [astro-ph.HE]}%
  \bibAnnoteFile{NoStop}{Gralla:2014yja}%
%%CITATION = ARXIV:1401.6159;%%
\bibitem{Philippov:2013tpa}%
  \BibitemOpen
  \bibfield{author}{%
  \bibinfo {author} {\bibfnamefont{A.}~\bibnamefont{Philippov}}\ and\ \bibinfo
  {author} {\bibfnamefont{A.}~\bibnamefont{Spitkovsky}}}%
   (\bibinfo {year} {2013}),\
  \Eprint{http://arxiv.org/abs/1312.4970}{arXiv:1312.4970 [astro-ph.HE]}%
  \bibAnnoteFile{NoStop}{Philippov:2013tpa}%
%%CITATION = ARXIV:1312.4970;%%
\bibitem{Mestel94}%
  \BibitemOpen
  \bibfield{author}{%
  \bibinfo {author} {\bibfnamefont{L.}~\bibnamefont{Mestel}}\ and\ \bibinfo
  {author} {\bibfnamefont{S.}~\bibnamefont{Shibata}},\ }%
  \bibfield{journal}{%
  \bibinfo {journal} {Mon.Not.Roy.Astron.Soc.}\ }%
  \textbf{\bibinfo {volume} {271}},\ \bibinfo {pages} {621} (\bibinfo {year}
  {1994})%
  \bibAnnoteFile{NoStop}{Mestel94}%
\bibitem{Contopoulos:1999ga}%
  \BibitemOpen
  \bibfield{author}{%
  \bibinfo {author} {\bibfnamefont{I.}~\bibnamefont{Contopoulos}}, \bibinfo
  {author} {\bibfnamefont{D.}~\bibnamefont{Kazanas}},\ and\ \bibinfo {author}
  {\bibfnamefont{C.}~\bibnamefont{Fendt}},\ }%
  \bibfield{journal}{%
  \Doi{10.1086/306652}{\bibinfo {journal} {Astrophys.J.}}\ }%
  \textbf{\bibinfo {volume} {511}},\ \bibinfo {pages} {351} (\bibinfo {year}
  {1999}),\
  \Eprint{http://arxiv.org/abs/astro-ph/9903049}{arXiv:astro-ph/9903049
  [astro-ph]}%
  \bibAnnoteFile{NoStop}{Contopoulos:1999ga}%
%%CITATION = ASTRO-PH/9903049;%%
\bibitem{Morrison:2004fp}%
  \BibitemOpen
  \bibfield{author}{%
  \bibinfo {author} {\bibfnamefont{I.~A.}\ \bibnamefont{Morrison}}, \bibinfo
  {author} {\bibfnamefont{T.~W.}\ \bibnamefont{Baumgarte}},\ and\ \bibinfo
  {author} {\bibfnamefont{S.~L.}\ \bibnamefont{Shapiro}},\ }%
  \bibfield{journal}{%
  \Doi{10.1086/421897}{\bibinfo {journal} {Astrophys.J.}}\ }%
  \textbf{\bibinfo {volume} {610}},\ \bibinfo {pages} {941} (\bibinfo {year}
  {2004}),\
  \Eprint{http://arxiv.org/abs/astro-ph/0401581}{arXiv:astro-ph/0401581
  [astro-ph]}%
  \bibAnnoteFile{NoStop}{Morrison:2004fp}%
%%CITATION = ASTRO-PH/0401581;%%
\bibitem{Palenzuela:2012my}%
  \BibitemOpen
  \bibfield{author}{%
  \bibinfo {author} {\bibfnamefont{C.}~\bibnamefont{Palenzuela}},\ }%
  \bibfield{journal}{%
  \Doi{10.1093/mnras/stt311}{\bibinfo {journal} {Mon. Not. R. Aston. Soc.}}\ }%
  \textbf{\bibinfo {volume} {431}},\ \bibinfo {pages} {, 1853} (\bibinfo {year}
  {2}),\ \Eprint{http://arxiv.org/abs/1212.0130}{arXiv:1212.0130
  [astro-ph.HE]}%
  \bibAnnoteFile{NoStop}{Palenzuela:2012my}%
%%CITATION = ARXIV:1212.0130;%%
\bibitem{Deutsch55}%
  \BibitemOpen
  \bibfield{author}{%
  \bibinfo {author} {\bibnamefont{{Deutsch Arnim J.,}.}},\ }%
  \bibfield{journal}{%
  \bibinfo {journal} {Annales d’Astrophysique}\ }%
  \textbf{\bibinfo {volume} {18}},\ \bibinfo {pages} {1} (\bibinfo {year}
  {1955})%
  \bibAnnoteFile{NoStop}{Deutsch55}%
\bibitem{Beskin90}%
  \BibitemOpen
  \bibfield{author}{%
  \bibinfo {author} {\bibnamefont{{Beskin, V. S.}.}},\ }%
  \bibfield{journal}{%
  \bibinfo {journal} {Soviet Astronomy Letters}\ }%
  \textbf{\bibinfo {volume} {16}},\ \bibinfo {pages} {286} (\bibinfo {year}
  {1990})%
  \bibAnnoteFile{NoStop}{Beskin90}%
\bibitem{Musmilov92}%
  \BibitemOpen
  \bibfield{author}{%
  \bibinfo {author} {\bibfnamefont{A.~I.}\ \bibnamefont{Muslimov},
  \bibfnamefont{A.~G.;~Tsygan}},\ }%
  \bibfield{journal}{%
  \bibinfo {journal} {Mon.Not.Roy.Astron.Soc.}\ }%
  \textbf{\bibinfo {volume} {255}},\ \bibinfo {pages} {61} (\bibinfo {year}
  {1992})%
  \bibAnnoteFile{NoStop}{Musmilov92}%
\bibitem{Petri:2014oza}%
  \BibitemOpen
  \bibfield{author}{%
  \bibinfo {author} {\bibfnamefont{J.}~\bibnamefont{P\'etri}}}%
   (\bibinfo {year} {2014}),\
  \Eprint{http://arxiv.org/abs/1401.1367}{arXiv:1401.1367 [astro-ph.HE]}%
  \bibAnnoteFile{NoStop}{Petri:2014oza}%
%%CITATION = ARXIV:1401.1367;%%
\bibitem{Rezzolla:2004hy}%
  \BibitemOpen
  \bibfield{author}{%
  \bibinfo {author} {\bibfnamefont{L.}~\bibnamefont{Rezzolla}}\ and\ \bibinfo
  {author} {\bibfnamefont{B.~J.}\ \bibnamefont{Ahmedov}},\ }%
  \bibfield{journal}{%
  \Doi{10.1111/j.1365-2966.2004.08006.x}{\bibinfo {journal}
  {Mon.Not.Roy.Astron.Soc.}}\ }%
  \textbf{\bibinfo {volume} {352}},\ \bibinfo {pages} {1161} (\bibinfo {year}
  {2004}),\ \Eprint{http://arxiv.org/abs/gr-qc/0406018}{arXiv:gr-qc/0406018
  [gr-qc]}%
  \bibAnnoteFile{NoStop}{Rezzolla:2004hy}%
%%CITATION = GR-QC/0406018;%%
\bibitem{Lehner:2011aa}%
  \BibitemOpen
  \bibfield{author}{%
  \bibinfo {author} {\bibfnamefont{L.}~\bibnamefont{Lehner}}, \bibinfo {author}
  {\bibfnamefont{C.}~\bibnamefont{Palenzuela}}, \bibinfo {author}
  {\bibfnamefont{S.~L.}\ \bibnamefont{Liebling}}, \bibinfo {author}
  {\bibfnamefont{C.}~\bibnamefont{Thompson}},\ and\ \bibinfo {author}
  {\bibfnamefont{C.}~\bibnamefont{Hanna}},\ }%
  \bibfield{journal}{%
  \Doi{10.1103/PhysRevD.86.104035}{\bibinfo {journal} {Phys.Rev.}}\ }%
  \textbf{\bibinfo {volume} {D86}},\ \bibinfo {pages} {104035} (\bibinfo {year}
  {2012}),\ \Eprint{http://arxiv.org/abs/1112.2622}{arXiv:1112.2622
  [astro-ph.HE]}%
  \bibAnnoteFile{NoStop}{Lehner:2011aa}%
%%CITATION = ARXIV:1112.2622;%%
\bibitem{Paschalidis:2013gma}%
  \BibitemOpen
  \bibfield{author}{%
  \bibinfo {author} {\bibfnamefont{V.}~\bibnamefont{Paschalidis}}\ and\
  \bibinfo {author} {\bibfnamefont{S.~L.}\ \bibnamefont{Shapiro}}}%
   (\bibinfo {year} {2013}),\
  \Eprint{http://arxiv.org/abs/1310.3274}{arXiv:1310.3274 [astro-ph.HE]}%
  \bibAnnoteFile{NoStop}{Paschalidis:2013gma}%
%%CITATION = ARXIV:1310.3274;%%
\bibitem{Wasserman83}%
  \BibitemOpen
  \bibfield{author}{%
  \bibinfo {author} {\bibfnamefont{I.}~\bibnamefont{Wasserman}}\ and\ \bibinfo
  {author} {\bibfnamefont{S.~L.}\ \bibnamefont{Shapiro}},\ }%
  \bibfield{journal}{%
  \Doi{10.1086/160745}{\bibinfo {journal} {Astrophys.J.}}\ }%
  \textbf{\bibinfo {volume} {265}},\ \bibinfo {pages} {1036} (\bibinfo {year}
  {1983})%
  \bibAnnoteFile{NoStop}{Wasserman83}%
%%CITATION = ASJOA,424,823;%%
\bibitem{Cook:1993qj}%
  \BibitemOpen
  \bibfield{author}{%
  \bibinfo {author} {\bibfnamefont{G.~B.}\ \bibnamefont{Cook}}, \bibinfo
  {author} {\bibfnamefont{S.~L.}\ \bibnamefont{Shapiro}},\ and\ \bibinfo
  {author} {\bibfnamefont{S.~A.}\ \bibnamefont{Teukolsky}}}%
  \bibfield{journal}{%
  \Doi{10.1086/173934}{\bibinfo {journal} {Astrophys.J.}}\ }%
  \textbf{\bibinfo {volume} {422}},\ \bibinfo {pages} {227}
   (\bibinfo {year} {1994})%
  \bibAnnoteFile{NoStop}{Cook:1993qj}%
%%CITATION = INSPIRE-360303;%%
\bibitem{Cook:1993qr}%
  \BibitemOpen
  \bibfield{author}{%
  \bibinfo {author} {\bibfnamefont{G.~B.}\ \bibnamefont{Cook}}, \bibinfo
  {author} {\bibfnamefont{S.~L.}\ \bibnamefont{Shapiro}},\ and\ \bibinfo
  {author} {\bibfnamefont{S.~A.}\ \bibnamefont{Teukolsky}},\ }%
  \bibfield{journal}{%
  \Doi{10.1086/173934}{\bibinfo {journal} {Astrophys.J.}}\ }%
  \textbf{\bibinfo {volume} {424}},\ \bibinfo {pages} {823} (\bibinfo {year}
  {1994})%
  \bibAnnoteFile{NoStop}{Cook:1993qr}%
%%CITATION = ASJOA,424,823;%%
\bibitem{Paschalidis:2013jsa}%
  \BibitemOpen
  \bibfield{author}{%
  \bibinfo {author} {\bibfnamefont{V.}~\bibnamefont{Paschalidis}}, \bibinfo
  {author} {\bibfnamefont{Z.~B.}\ \bibnamefont{Etienne}},\ and\ \bibinfo
  {author} {\bibfnamefont{S.~L.}\ \bibnamefont{Shapiro}},\ }%
  \bibfield{journal}{%
  \Doi{10.1103/PhysRevD.88.021504}{\bibinfo {journal} {Phys.Rev.}}\ }%
  \textbf{\bibinfo {volume} {D88}},\ \bibinfo {pages} {021504} (\bibinfo {year}
  {2013}),\ \Eprint{http://arxiv.org/abs/1304.1805}{arXiv:1304.1805
  [astro-ph.HE]}%
  \bibAnnoteFile{NoStop}{Paschalidis:2013jsa}%
%%CITATION = ARXIV:1304.1805;%%
\bibitem{Blandford02}%
  \BibitemOpen
  \bibfield{author}{%
  \bibinfo {author} {\bibnamefont{{R. D. Blandford and R. L. Znajek}.}},\ }%
  \bibfield{journal}{%
  \bibinfo {journal} {Mon. Not. R. Astr. Soc.}\ }%
  \textbf{\bibinfo {volume} {179}},\ \bibinfo {pages} {433} (\bibinfo {year}
  {1977})%
  \bibAnnoteFile{NoStop}{Blandford02}%
\bibitem{McKinney:2006sc}%
  \BibitemOpen
  \bibfield{author}{%
  \bibinfo {author} {\bibfnamefont{J.~C.}\ \bibnamefont{McKinney}},\ }%
  \bibfield{journal}{%
  \Doi{10.1111/j.1365-2966.2006.10087.x}{\bibinfo {journal}
  {Mon.Not.Roy.Astron.Soc.}}\ }%
  \textbf{\bibinfo {volume} {367}},\ \bibinfo {pages} {1797} (\bibinfo {year}
  {2006}),\
  \Eprint{http://arxiv.org/abs/astro-ph/0601410}{arXiv:astro-ph/0601410
  [astro-ph]}%
  \bibAnnoteFile{NoStop}{McKinney:2006sc}%
%%CITATION = ASTRO-PH/0601410;%%
\bibitem{Duez:2005sf}%
  \BibitemOpen
  \bibfield{author}{%
  \bibinfo {author} {\bibfnamefont{M.~D.}\ \bibnamefont{Duez}}, \bibinfo
  {author} {\bibfnamefont{Y.~T.}\ \bibnamefont{Liu}}, \bibinfo {author}
  {\bibfnamefont{S.~L.}\ \bibnamefont{Shapiro}},\ and\ \bibinfo {author}
  {\bibfnamefont{B.~C.}\ \bibnamefont{Stephens}},\ }%
  \bibfield{journal}{%
  \Doi{10.1103/PhysRevD.72.024028}{\bibinfo {journal} {Phys.Rev.}}\ }%
  \textbf{\bibinfo {volume} {D72}},\ \bibinfo {pages} {024028} (\bibinfo {year}
  {2005}),\
  \Eprint{http://arxiv.org/abs/astro-ph/0503420}{arXiv:astro-ph/0503420
  [astro-ph]}%
  \bibAnnoteFile{NoStop}{Duez:2005sf}%
%%CITATION = ASTRO-PH/0503420;%%
\bibitem{Etienne:2010ui}%
  \BibitemOpen
  \bibfield{author}{%
  \bibinfo {author} {\bibfnamefont{Z.~B.}\ \bibnamefont{Etienne}}, \bibinfo
  {author} {\bibfnamefont{Y.~T.}\ \bibnamefont{Liu}},\ and\ \bibinfo {author}
  {\bibfnamefont{S.~L.}\ \bibnamefont{Shapiro}},\ }%
  \bibfield{journal}{%
  \Doi{10.1103/PhysRevD.82.084031}{\bibinfo {journal} {Phys.Rev.}}\ }%
  \textbf{\bibinfo {volume} {D82}},\ \bibinfo {pages} {084031} (\bibinfo {year}
  {2010}),\ \Eprint{http://arxiv.org/abs/1007.2848}{arXiv:1007.2848
  [astro-ph.HE]}%
  \bibAnnoteFile{NoStop}{Etienne:2010ui}%
%%CITATION = ARXIV:1007.2848;%%
\bibitem{Etienne:2011re}%
  \BibitemOpen
  \bibfield{author}{%
  \bibinfo {author} {\bibfnamefont{Z.~B.}\ \bibnamefont{Etienne}}, \bibinfo
  {author} {\bibfnamefont{V.}~\bibnamefont{Paschalidis}}, \bibinfo {author}
  {\bibfnamefont{Y.~T.}\ \bibnamefont{Liu}},\ and\ \bibinfo {author}
  {\bibfnamefont{S.~L.}\ \bibnamefont{Shapiro}},\ }%
  \bibfield{journal}{%
  \Doi{10.1103/PhysRevD.85.024013}{\bibinfo {journal} {Phys.Rev.}}\ }%
  \textbf{\bibinfo {volume} {D85}},\ \bibinfo {pages} {024013} (\bibinfo {year}
  {2012}),\ \Eprint{http://arxiv.org/abs/1110.4633}{arXiv:1110.4633
  [astro-ph.HE]}%
  \bibAnnoteFile{NoStop}{Etienne:2011re}%
%%CITATION = ARXIV:1110.4633;%%
\bibitem{PhysRevLett.109.221102}%
  \BibitemOpen
  \bibfield{author}{%
  \bibinfo {author} {\bibfnamefont{B.~D.}\ \bibnamefont{Farris}}, \bibinfo
  {author} {\bibfnamefont{R.}~\bibnamefont{Gold}}, \bibinfo {author}
  {\bibfnamefont{V.}~\bibnamefont{Paschalidis}}, \bibinfo {author}
  {\bibfnamefont{Z.~B.}\ \bibnamefont{Etienne}},\ and\ \bibinfo {author}
  {\bibfnamefont{S.~L.}\ \bibnamefont{Shapiro}},\ }%
  \bibfield{journal}{%
  \Doi{10.1103/PhysRevLett.109.221102}{\bibinfo {journal} {Phys. Rev. Lett.}}\
  }%
  \textbf{\bibinfo {volume} {109}},\ \bibinfo {pages} {221102} (\bibinfo
  {month} {Nov}\ \bibinfo {year} {2012}),\
  \url{http://link.aps.org/doi/10.1103/PhysRevLett.109.221102}%
  \bibAnnoteFile{NoStop}{PhysRevLett.109.221102}%
\bibitem{Gold:2013zma}%
  \BibitemOpen
  \bibfield{author}{%
  \bibinfo {author} {\bibfnamefont{R.}~\bibnamefont{Gold}}, \bibinfo {author}
  {\bibfnamefont{V.}~\bibnamefont{Paschalidis}}, \bibinfo {author}
  {\bibfnamefont{Z.~B.}\ \bibnamefont{Etienne}}, \bibinfo {author}
  {\bibfnamefont{S.~L.}\ \bibnamefont{Shapiro}},\ and\ \bibinfo {author}
  {\bibfnamefont{H.~P.}\ \bibnamefont{Pfeiffer}}}%
   (\bibinfo {year} {2013}),\
  \Eprint{http://arxiv.org/abs/1312.0600}{arXiv:1312.0600 [astro-ph.HE]}%
  \bibAnnoteFile{NoStop}{Gold:2013zma}%
%%CITATION = ARXIV:1312.0600;%%
\bibitem{Colella84}%
  \BibitemOpen
  \bibfield{author}{%
  \bibinfo {author} {\bibfnamefont{P.}~\bibnamefont{Colella}}\ and\ \bibinfo
  {author} {\bibfnamefont{P.~R.}\ \bibnamefont{Woodward}},\ }%
  \bibfield{journal}{%
  \bibinfo {journal} {J. Comput. Phys.}\ }%
  \textbf{\bibinfo {volume} {54}},\ \bibinfo {pages} {174} (\bibinfo {year}
  {1984})%
  \bibAnnoteFile{NoStop}{Colella84}%
\bibitem{Harten83}%
  \BibitemOpen
  \bibfield{author}{%
  \bibinfo {author} {\bibfnamefont{A.}~\bibnamefont{Harten}}, \bibinfo {author}
  {\bibfnamefont{P.~D.}\ \bibnamefont{Lax}},\ and\ \bibinfo {author}
  {\bibfnamefont{B.}~\bibnamefont{van Leer}},\ }%
  \bibfield{journal}{%
  \bibinfo {journal} {SIAM Rev.}\ }%
  \textbf{\bibinfo {volume} {25}},\ \bibinfo {pages} {35} (\bibinfo {year}
  {1983})%
  \bibAnnoteFile{NoStop}{Harten83}%
\bibitem{Akmal98}%
  \BibitemOpen
  \bibfield{author}{%
  \bibinfo {author} {\bibfnamefont{A.}~\bibnamefont{Akmal}}, \bibinfo {author}
  {\bibfnamefont{V.~R.}\ \bibnamefont{Pandharipande}},\ and\ \bibinfo {author}
  {\bibfnamefont{D.~G.}\ \bibnamefont{Ravenhall}},\ }%
  \bibfield{journal}{%
  \Doi{10.1103/PhysRevC.58.1804}{\bibinfo {journal} {Phys. Rev. C}}\ }%
  \textbf{\bibinfo {volume} {58}},\ \bibinfo {pages} {1804} (\bibinfo {month}
  {Sep}\ \bibinfo {year} {1998}),\
  \url{http://link.aps.org/doi/10.1103/PhysRevC.58.1804}%
  \bibAnnoteFile{NoStop}{Akmal98}%
\bibitem{Lackey:2011vz}%
  \BibitemOpen
  \bibfield{author}{%
  \bibinfo {author} {\bibfnamefont{B.~D.}\ \bibnamefont{Lackey}}, \bibinfo
  {author} {\bibfnamefont{K.}~\bibnamefont{Kyutoku}}, \bibinfo {author}
  {\bibfnamefont{M.}~\bibnamefont{Shibata}}, \bibinfo {author}
  {\bibfnamefont{P.~R.}\ \bibnamefont{Brady}},\ and\ \bibinfo {author}
  {\bibfnamefont{J.~L.}\ \bibnamefont{Friedman}},\ }%
  \bibfield{journal}{%
  \Doi{10.1103/PhysRevD.85.044061}{\bibinfo {journal} {Phys.Rev.}}\ }%
  \textbf{\bibinfo {volume} {D85}},\ \bibinfo {pages} {044061} (\bibinfo {year}
  {2012}),\ \Eprint{http://arxiv.org/abs/1109.3402}{arXiv:1109.3402
  [astro-ph.HE]}%
  \bibAnnoteFile{NoStop}{Lackey:2011vz}%
%%CITATION = ARXIV:1109.3402;%%
\bibitem{Bauswein:2011tp}%
  \BibitemOpen
  \bibfield{author}{%
  \bibinfo {author} {\bibfnamefont{A.}~\bibnamefont{Bauswein}}\ and\ \bibinfo
  {author} {\bibfnamefont{H.-T.}\ \bibnamefont{Janka}},\ }%
  \bibfield{journal}{%
  \Doi{10.1103/PhysRevLett.108.011101}{\bibinfo {journal} {Phys.Rev.Lett.}}\ }%
  \textbf{\bibinfo {volume} {108}},\ \bibinfo {pages} {011101} (\bibinfo {year}
  {2012}),\ \Eprint{http://arxiv.org/abs/1106.1616}{arXiv:1106.1616
  [astro-ph.SR]}%
  \bibAnnoteFile{NoStop}{Bauswein:2011tp}%
%%CITATION = ARXIV:1106.1616;%%
\bibitem{Paschalidis:2012ff}%
  \BibitemOpen
  \bibfield{author}{%
  \bibinfo {author} {\bibfnamefont{V.}~\bibnamefont{Paschalidis}}, \bibinfo
  {author} {\bibfnamefont{Z.~B.}\ \bibnamefont{Etienne}},\ and\ \bibinfo
  {author} {\bibfnamefont{S.~L.}\ \bibnamefont{Shapiro}},\ }%
  \bibfield{journal}{%
  \Doi{10.1103/PhysRevD.86.064032}{\bibinfo {journal} {Phys.Rev.}}\ }%
  \textbf{\bibinfo {volume} {D86}},\ \bibinfo {pages} {064032} (\bibinfo {year}
  {2012}),\ \Eprint{http://arxiv.org/abs/1208.5487}{arXiv:1208.5487
  [astro-ph.HE]}%
  \bibAnnoteFile{NoStop}{Paschalidis:2012ff}%
%%CITATION = ARXIV:1208.5487;%%
\end{thebibliography}%
%%%%%%%%%%%%%%%%%%%%%%%%%%%%%%%%%%%%%%%%  
\end{document}